\documentclass{article}
\newlength{\abstractwidth}
\usepackage{cancel}
\usepackage{hyperref}
\usepackage{color}
\usepackage{graphicx}
\usepackage{tikz}
\usepackage{mathtools}
\usepackage{float}

\usepackage{cancel}
\usepackage{hyperref}

\usepackage{amsmath}
\usepackage{amssymb}
\usepackage{latexsym}
 \usepackage{setspace}
 
 \usepackage{caption}
\usepackage{subcaption}

\numberwithin{equation}{section}

\renewcommand{\thefootnote}{\fnsymbol{footnote}}
\renewcommand{\thanks}[1]{\footnote{#1}}
\newcommand{\starttext}{
\setcounter{footnote}{0}
\renewcommand{\thefootnote}{\arabic{footnote}}}
\newcommand{\bea}{\begin{eqnarray}}
\newcommand{\eea}{\end{eqnarray}}
\newcommand{\be}{\begin{eqnarray}}
\newcommand{\ee}{\end{eqnarray}}

\def\ie{\begin{equation}\begin{aligned}}
\def\fe{\end{aligned}\end{equation}}
\def\half{{\scriptstyle \frac 12}}

\def\sevenh{{\scriptstyle \frac 72}}
\def\threeh{{\scriptstyle \frac 32}}
\def\fiveh{{\scriptstyle \frac 52}}

\def\ie{\begin{equation}\begin{aligned}}
\def\fe{\end{aligned}\end{equation}}

\def\cC{{\cal C}}

\def\cN{{\cal N}}
\def\cO{{\cal O}}

\def\cQ{{\cal Q}}

\def\Z{{\mathbb Z}}

\def\RR{{\mathbb R}}

\def\nn{\nonumber}

\def\Im{{\rm Im \,}}
\def\tr{{\rm tr}}

\def\C{\cC}

\def\flux{{\tilde{N}}_{G_N}}
\def\fluxSU{{\tilde{N}}_{SU(N)}}
\def\fluxSO{{\tilde{N}}_{SO(n)}}
\def\fluxUSp{{\tilde{N}}_{USp(n)}}

\begin{document}

\starttext

\setcounter{footnote}{0}

\begin{flushright}
%\scriptsize 
{\small QMUL-PH-22-04}
\end{flushright}

\vskip 0.3in

\begin{center}

\centerline{\large \bf Exact results for duality-covariant integrated correlators} 

\centerline{\large \bf  in $\cN=4$  SYM with general classical gauge groups}

\vskip 0.2in

{ Daniele Dorigoni$^{(a)}$, Michael B. Green$^{(b),(c)}$  and Congkao Wen$^{(c)}$} 
   
\vskip 0.15in

{\small ($a$) Centre for Particle Theory \& Department of Mathematical Sciences, 
}\\
\small{Durham University, Lower Mountjoy, Stockton Road, Durham DH1 3LE, UK}

\vskip 0.1in

{ \small ($b$) Department of Applied Mathematics and Theoretical Physics }\\
{\small  Wilberforce Road, Cambridge CB3 0WA, UK}

\vskip 0.1in

{\small  ($c$) Centre for Theoretical Physics, Department of Physics and Astronomy,  }\\ 
{\small Queen Mary University of London,  London, E1 4NS, UK}

\vskip 0.5in

\begin{abstract}
\vskip 0.1in
We present exact expressions for certain integrated correlators of four superconformal primary operators in the stress tensor multiplet of $\cN=4$ supersymmetric Yang--Mills (SYM) theory with classical gauge group, $G_N$ $= SO(2N)$, $SO(2N+1)$, $USp(2N)$. These integrated correlators are expressed as two-dimensional lattice sums by considering derivatives of the localised partition functions, generalising the expression obtained for $SU(N)$ {gauge group}  in our previous works.   
 These expressions are manifestly covariant under Goddard-Nuyts-Olive duality.   
The integrated correlators can also be formally written as infinite sums of non-holomorphic Eisenstein series with integer indices and rational coefficients.  
  Furthermore, the action of the hyperbolic Laplace operator with respect to the complex coupling $\tau=\theta/(2\pi) + 4\pi i  /g^2_{_{YM}}$ on any integrated correlator for gauge group $G_N$ relates it to  a linear combination of correlators with gauge groups $G_{N+1}$, $G_N$ and $G_{N-1}$. These ``Laplace-difference equations'' determine the expressions of integrated correlators for all  classical gauge groups for any value of  $N$ in terms of the correlator for the gauge group $SU(2)$.
The perturbation expansions of these integrated  correlators for any finite value of $N$ agree with properties obtained from perturbative Yang--Mills quantum field theory, together with various multi-instanton calculations which are also shown to agree with those determined by supersymmetric localisation.  The coefficients of terms in the large-$N$ expansion are sums of non-holomorphic Eisenstein series with half-integer indices, which  extend recent results and make contact with low order terms in the low energy expansion of type IIB superstring theory in an $AdS_5\times S^5/\mathbb{Z}_2$ background.
\end{abstract}                                            
\end{center}

\newpage

\tableofcontents
\newpage

\section{Introduction and outline}

In \cite{Dorigoni:2021bvj,Dorigoni:2021guq} an integrated correlator of four superconformal primary operators in the stress tensor multiplet of $\cN=4$ supersymmetric $SU(N)$ Yang--Mills (SYM)  theory was expressed as a two-dimensional lattice sum that is manifestly invariant under $SL(2,\Z)$ Montonen-Olive  duality and is valid for all values of $N$ and the coupling constant $\tau=\tau_1+i \tau_2 = \theta/(2\pi) + i 4\pi/g_{_{YM}}^2$ in the upper-half plane $\tau_2>0$. \footnote{The action of $SL(2, \Z)$ is:  
$\tau  \underset {SL(2,\Z)} \to  (a\tau+b)/(c\tau+d)$, where  $a,b,c,d\in \Z$ and $ad-bc=1$. }
 This correlator  was originally defined in  \cite{Binder:2019jwn} in terms of derivatives acting on the localised partition function of the $\cN=2^*$ SYM theory on $S^4$ \cite{Pestun:2007rz}, which can be expressed as a mass deformation of the $\cN=4$ theory.   The $\cN=4$ integrated correlator results from the $m\to 0$ limit (where $m$ is the hypermultiplet mass). In this paper, we will consider an integrated correlator for $\cN=4$ SYM with any classical gauge group  $G_N$ $=SU(N)$,  $SO(2N)$, $SO(2N+1)$,  $USp(2N)$, which is given by 
  \bea \label{integratedF1}
\C_{G_N} (\tau,\bar\tau) = \left.  {1\over 4} \, { \Delta_{\tau}\partial_m^2 \log Z_{G_N}}(m, \tau, \bar{\tau})    \right |_{m=0}   \, ,
\label{firstmeasure}
\eea
where $Z_{G_N} (m,\tau,\bar\tau)$ is the partition function of $\cN=2^*$ SYM on $S^4$ with a gauge group $G_N$, $\C_{G_N} (\tau,\bar\tau)$ denotes the integrated four-point correlator and $\Delta_\tau = \tau_2^2(\partial_{\tau_1}^2+\partial_{\tau_2}^2)$ is the laplacian on the hyperbolic plane. The expression \eqref{firstmeasure} was shown in   \cite{Binder:2019jwn} to define  a four-point correlator integrated over the positions of the operators with a specific measure that has the following schematic form 
 \bea
 \int \prod_{i=1}^4 dx_i \, \mu(x_1,\dots,x_4)\,\langle \cO_2(x_1) \dots \cO_2(x_4)\rangle\, ,
 \label{intcorr}
 \eea
 where $\cO_2(x)$ denotes the superconformal primary operator in the stress tensor supermultiplet, which is in the ${\bf 20^\prime}$ of the $SU(4)$ R-symmetry group  and $ \mu(x_1,\dots,x_4)$ is a measure factor. The precise expression for \eqref{intcorr} is discussed in \cite{Binder:2019jwn}  and later references. 
 Some properties of the large-$N$ expansion of $\C_{SU(N)}(\tau,\bar\tau)$  were considered in   \cite{Chester:2019pvm, Chester:2020dja,Chester:2019jas}.\footnote{In these references the correlator was denoted $\mathcal{G}_N(\tau,\bar\tau)$.}

A second integrated correlator of the form \eqref{intcorr} but with a different integration measure was introduced in  \cite{Chester:2019jas},  and is proportional to
 $\partial_m^4 \log Z_{SU(N)}(m, \tau, \bar{\tau})    \big |_{m=0}$.  Some properties of its large-$N$ expansion were  elucidated in \cite{Chester:2020vyz} and more recently in \cite{Collier:2022emf}.  We will not consider this integrated correlator in this paper.
  
\subsection{The main results}
\label{sec:main}

In this paper we will consider the extension of the $SU(N)$ results of \cite{Dorigoni:2021bvj,Dorigoni:2021guq} to the other classical Lie groups,  $SO(2N)$, $SO(2N+1)$, and $USp(2N)$.  Some aspects of the perturbative expansions of the integrated correlators for these groups,  and their large-$N$  expansions in the  't Hooft limit were considered in \cite{Alday:2021vfb} starting from the localised partition function of $\cN=2^*$ SYM described in \cite{Pestun:2007rz}.
Our analysis will include the non-perturbative instanton contributions, leading to expressions for the integrated correlators for $\cN=4$ SYM with any classical gauge group that take the form of two-dimensional lattice sums\footnote{As we will clarify later, the $SO(3)$ case is an exception, and in that case the integrated correlator is $\C_{SO(3)}(\frac{\tau}{2},\frac{\bar\tau}{2})$ (rather than $\C_{SO(3)}(\tau,\bar\tau)$),  which agrees with the result of supersymmetric localisation.}
\bea
\C_{G_N}(\tau,\bar\tau) =  \sum_{(m,n)\in \Z^2} \int_0^\infty dt\left(B^1_{G_N}(t) e^{-t \pi\frac{ |m+n\tau|^2}{\tau_2}}+B_{G_N}^2 (t) e^{-t\pi \frac{ |m+ 2 n \tau|^2}{ 2 \tau_2}}\right)\,.
\label{mainres}
\eea
The rational functions $B^1_{G_N}(t)$ and $B^2_{G_N}(t)$ will be defined in detail later. Here we note that in the simply-laced cases, i.e. $SU(N)$ and $SO(2N)$, we have $B_{G_N}^2 (t)=0$ so that we may drop the superscript and denote $B_{G_N}^1(t) = B_{G_N}(t)$. In these cases the expression is manifestly  invariant under $SL(2,\Z)$, which is generated by the transformations $S$ and $T$ where $T: \tau \to \tau+1$ and $S: \tau\to - 1/\tau $. This was originally suggested by Montonen and Olive \cite{Montonen:1977sn, Witten:1978mh,  Osborn:1979tq} following the observations by Goddard, Nuyts and Olive (GNO) concerning  the relation between electric charge and magnetic monopole weight lattices in gauge field theories \cite{Goddard:1976qe}.  
 
In the non simply-laced cases, i.e. $SO(2N+1)$ and  $USp(2N)$, the expression \eqref{mainres} is invariant under $\Gamma_0(2) \subset SL(2,\Z)$.\footnote{An element $\gamma \!=\! \begin{psmallmatrix}a & b\\c & d\end{psmallmatrix} \!\in\! SL(2,\mathbb{Z})$ belongs to the congruence subgroup $\Gamma_0(2)$ if $c = 0 \,{\rm mod}\,2$.} This is the group generated by $T$ and ${\hat S}\, T\, {\hat S}$, where $\hat S: \tau\to -1/(2\tau)$, $T: \tau\to \tau+1$.  The action of  $\hat S$ does not leave \eqref{mainres} invariant but rather interchanges the two terms. However, we will see that 
\bea
B^1_{SO(2N+1)} (t)= B^2_{USp(2N)}  (t)\,,\qquad\quad B^1_{USp(2N)} (t)  = B^2_{SO(2N+1) }(t) \, ,
\label{nonsimp}
\eea
so that $\hat S$ acts as a GNO (or Langlands) duality transformation \cite{Girardello:1995gf, Dorey:1996hx, Kapustin:2006pk}, which relates $\C_{SO(2N+1)}$ with $\C_{USp(2N)}$. Since we are only concerned with correlation functions of local operators, effectively GNO duality acts at the level of Lie algebras rather than Lie groups. The global versions of GNO duality are briefly reviewed in appendix~\ref{secGNO}.

Detailed discussion of these results will be given in  later sections but here we note the following general points:

$\bullet$
As in the $SU(N)$ case considered in  \cite{Dorigoni:2021bvj,Dorigoni:2021guq} the functions $B_{G_N}^i(t)$ ($i=1,2$)  
satisfy inversion conditions  
\bea
B_{G_N}^i  (t) = t^{-1}\,B_{G_N}^i (t^{-1})\, ,
\label{invert}
\eea
and integration conditions 
\begin{align}
 \int_0^\infty  dt \,B_{SU(N)} (t)  &= \frac{N(N-1)}{8} \,,  \nn\\
 \int_0^\infty  dt\,B_{SO(2N)} (t) &=  \int_0^\infty dt \, B_{SO(2N+1)}^1 (t)   = \int_0^\infty dt \, B_{USp(2N)}^2 (t) = \frac{N(N-1)}{4} \,,\nn\\
    \int_0^\infty dt\,  B_{SO(2N+1)}^2 (t)   &= \int_0^\infty dt\,  B_{USp(2N)}^1 (t) = \frac{N}{4} \,, 
\label{intcon}
\end{align}
as well as 
\bea
 \int_0^\infty \frac{dt}{\sqrt t}\, B^i_{G_N} (t)=0\,.
  \label{zeroint}
  \eea

$\bullet$  The integrated correlator \eqref{mainres} can be expressed as a formal expansion of the form
\bea
\C_{G_N}(\tau,\bar\tau)  =-b_{G_ N}(0) +\sum_{s=2}^\infty  \left[ b_{G_N}^1 (s)\, E(s;\tau,\bar\tau)  + b_{G_N}^2 (s)\,E(s;2\tau,2\bar\tau)\right]  \,,
\label{eisenform}
\eea
where $E(s;\tau,\bar\tau)$ is a non-holomorphic (or real analytic) Eisenstein series with $s\in \mathbb{N}$ (in our convention $E(0;\tau,\bar\tau)=-1$).
The coefficients $b^1_{G_N}(s)$ and $b^2_{G_N}(s)$ are rational numbers that are determined by the expansion of  $B^i_{G_N}(t)$ in the form
\bea
B_{G_N}^i (t) = \sum_{s=2}^\infty \frac{b_{G_N}^i (s)}{\Gamma(s)}\,\, t^{s-1}\,, \qquad \qquad i=1,2 \, ,
\label{Bexpand}
\eea
and $b_{G_N} (0) = b_{G_N}^1(0) + b_{G_N}^2(0)$ (since $b_{G_N}^2(s) = 0$ for $G_N= SU(N)$ and $SO(2N)$, in these cases we will drop the superscript and write $b_{G_N}^1(s) = b_{G_N}(s)$).

$\bullet$ It was  pointed out in \cite{Collier:2022emf}, in the $SU(N)$ case that the formal expression \eqref{eisenform} can be written in a manifestly convergent manner using the conventional spectral representation for a modular invariant function. Similarly, \eqref{eisenform} (which is a $\Gamma_0(2)$ invariant expression  in the $SO(2N+1)$ and $USp(2N)$ cases) has the form,
\begin{equation}
\C_{G_N}(\tau,\bar\tau) = -2 b_{G_N}(0) + \frac{1}{2\pi i } \int_{\frac{1}{2}-i\infty}^{\frac{1}{2}+i\infty} ds\,\frac{\pi (-1)^s}{\sin \pi s} \left[ b^1_{G_N}(s) E(s;\tau,\bar{\tau}) + b^2_{G_N}(s)E_s(s;2\tau,2\bar{\tau})\right]\,.
\end{equation}
 In \cite{Collier:2022emf} it was shown that the constant $-2 b_{G_N}(0)$ is equal to the ensemble average $\langle\C_{G_N}\rangle$, i.e. the integral of $\C_{G_N}(\tau,\bar{\tau})$ over the $\mathcal{N}=4$ conformal manifold, with respect to the Zamolodchikov metric. 

$\bullet$  The expressions \eqref{mainres} and \eqref{eisenform}  transform covariantly under GNO duality.  In the simply-laced cases, $SU(N)$ and $SO(2N)$, the coefficients $b_{G_N}^2 (s)$ vanish. Since $E(s;\tau,\bar\tau)$ is a modular function, the integrated correlators in these cases are invariant under $SL(2,\Z)$.

$\bullet$   In the non simply-laced cases, $SO(2N+1)$ and $USp(2N)$, it follows from  \eqref{nonsimp} and \eqref{Bexpand} that $\C_{G_N}(\tau,\bar\tau)$ given by \eqref{mainres} is invariant under the $\Gamma_0(2)$ subgroup of $SL(2,\Z)$ that is generated by the transformations $T$ and $\hat{S} T \hat{S}$.  The action of $\hat S$ on $\C_{G_N}(\tau,\bar\tau)$ effectively interchanges $b_{G_N}^1 (s)$ and $ b_{G_N}^2 (s)$ since
\bea
E(s;\tau,\bar\tau) \underset{\hat S}{\to} E\Big(s;-\frac{1}{2\tau},-\frac{1}{2\bar\tau}\Big) = E(s;2\tau,2\bar\tau)\,.
\label{shat onE}
\eea
This interchanges the integrated correlators for the $SO(2N+1)$ and $USp(2N)$ cases, and is interpreted as a GNO duality transformation. 

$\bullet$ The integrated correlators also satisfy Laplace-difference equations that generalise the equation satisfied in the $SU(N)$ case in  \cite{Dorigoni:2021bvj,Dorigoni:2021guq}.  These take the schematic form 
\begin{align}
\Delta_\tau \C_{G_N} &  -2 c_{G_N}   \Big[\C_{G_{N+1}} -2 \,\C_{G_N}+\C_{G_{N-1}}\Big] \cr
&+d_{G_{N+1}} \C_{SU(2N-1)}+d_{G_{N}} \C_{SU(2N)}+  d_{G_{N-1}} \C_{SU(2N+1)} =0\, ,
\label{lapdiffGN}
\end{align}
for $G_N = SO(2N)$, $SO(2N+1)$, $USp(2N)$, and 
where $c_{G_N}$ is the central charge. The precise values for the coefficients $d_{G_{N+1}}, d_{G_{N}}, d_{G_{N-1}} $ will be given later  for each gauge group. These equations also display the anticipated covariance under GNO duality. 
 
$\bullet$  The Laplace-difference equation for $SO(2N)$ is mapped into the Laplace-difference equation for $USp(2N)$ under the transformation $N\to -N$, together with $\tau \to -2\tau$. We will furthermore see that the  perturbative expansions of the integrated correlators confirm the identification of  $\C_{SO(-2N)} (- \tau,-\bar\tau)  $ and  $\C_{USp(2N)} (2\tau, 2\bar\tau)$.

$\bullet$   The large-$N$ expansion is naturally expressed as an expansion in inverse half-integer powers of the Ramond--Ramond (RR) five-form flux,

\subsection{Outline} 

In section~\ref{intgencorr} we will present some properties of the integrated correlator, $\C_{G_N}$,  defined in \eqref{firstmeasure} in terms of derivatives of the partition function of $\cN=2^*$ SYM on $S^4$ in the $m\to 0$ limit.  These results are based on methods outlined in appendix~\ref{pertlocal}, which   includes a brief summary of the perturbative structure of integrated correlators given in  \cite{Alday:2021vfb}, and an overview of instanton calculations based on the Nekrasov partition function \cite{Nekrasov:2002qd}  generalisied to arbitrary classical gauge groups  \cite{Billo:2015pjb, Billo:2015jyt}.    The   perturbation expansions for $\C_{G_N}(\tau_2)$ with finite $N$ are  presented in section~\ref{pertexpn}.  The  expansions for $\C_{SO(2N)}$, $\C_{SO(2N+1)}$ and $\C_{USp(2N)}$ generalise the expansion of $\C_{SU(N)}$ and display a number of interesting features, such as  the  equality of the $SO(2N)$  and $USp(-2N)$  integrated correlators when $g_{_{YM}}^2 \to -  2g_{_{YM}}^2$.   Furthermore, when expressed in terms of appropriate expansion parameters  all three of these integrated correlators have identical planar contributions (where the definition of 'planar' is dependent on the gauge group).   Non-planar contributions begin at $O\left((g_{_{YM}}^2)^4 \right)$.  The instanton contributions to $\C_{G_N}$  are discussed in section~\ref{sec:yminst} based on the formalism described in appendix~\ref{instcont}. The explicit form of these instanton contributions to $\C_{G_N}$  is difficult to extract from the localised partition function for general instanton number.  However, we have determined the exact expressions for the one-instanton sector, and to a certain extent the two- and three-instanton sectors.  

In  section \ref{sec:lapdiff} we will demonstrate that the perturbative parts of the  integrated correlators satisfy `Laplace-difference' equations  that have a form illustrated in \eqref{lapdiffGN}, which imply powerful constraints on their structure.  
 By studying various examples of  these equations we  are led in section~\ref{sec:ansatz}  to    conjecture that the fully non-perturbative expression for an integrated correlator  $\C_{G_N}(\tau, \bar \tau)$  can be expressed as the two-dimensional lattice sum in \eqref{mainres}, which is formally equivalent to the infinite sum of non-holomorphic Eisenstein series of integer index in  \eqref{eisenform}. These expressions transform in a manifestly covariant fashion under GNO duality.  They also  
contain an infinite number of Yang--Mills instanton contributions with precisely specified properties, which we will demonstrate agree with the instanton contributions to the localised correlators obtained in section~\ref{sec:yminst}.  The arguments that motivate the Laplace-difference equations are presented in appendix~\ref{sec:laplace-dif}.

 In section~\ref{sec:largen} we will consider the large-$N$ expansion of $\C_{G_N}(\tau,\bar\tau)$ in various limits of the Yang--Mills coupling.  In both the weakly-coupled and strongly-coupled  't Hooft limits considered in section~\ref{thooftn} the instanton contributions are suppressed exponentially in $N$ and only the perturbative terms contribute.  As we will show, if we introduce suitable expansion parameters  the perturbative expansions for different gauge groups are closely related.    The  definitions of these parameters, which are generalisations of the parameters $N$ and $g_{_{YM}}^2N$ for the $SU(N)$ case, that are suited to the large-$N$  weak-coupling expansion are not generally the same as the parameters suited to the large-$N$ strong-coupling expansion.  In section~\ref{fixedg}, we consider the large-$N$ limit with fixed $g_{_{YM}}^2$, where the instanton contribution is crucial for exhibiting manifest invariance under GNO duality. The expressions for the integrated correlators, which are obtained by solving Laplace-difference equations, take their most compact form when expanded in inverse (half-integral) powers of Ramond--Ramond  five-form flux $\flux$. The powers of $1/\flux$ correspond to powers of $\alpha^\prime$ in the low energy expansion of the holographic dual string theory and beautifully match the expected string theory structure.
               
We will end in section~\ref{sec:discussion} with a discussion of these results and of  possible future directions.     
    
\section{Integrated correlators for general  classical Lie groups} 
\label{intgencorr}

In this section we will determine properties of the perturbative and instantonic contributions to the integrated correlators based on  supersymmetric localisation.  The perturbative terms are contained in the zero Fourier mode with respect to $\tau_1$ whereas the non-perturbative terms correspond to the sum over instantons with instanton number $k\ne0$.  In other words, we can  express the correlator as a Fourier series,
\bea
\C_{G_N} (\tau,\bar\tau) = \C_{G_N}^{(0)} (\tau_2) + \sum_{k=1}^\infty  \left(e^{2\pi i k \tau}\, \C_{G_N}^{(k)}(\tau_2) +e^{-2\pi i k \bar\tau}\, \C_{G_N}^{(-k)}(\tau_2) \right) \, ,
\label{fmodes}
\eea
where the $k=0$ term is the perturbative contribution,
\bea
\C_{G_N}^{pert} (\tau_2) :=\C_{G_N}^{(0)} (\tau_2)  \,,
\label{propseries}
\eea
and the $k\ne 0$ terms are the instanton and anti-instanton contributions,
\bea
\C_{G_N}^{inst} (\tau,\bar\tau)\  := \sum_{k=1}^\infty  \left(e^{2\pi i k \tau}\, \C_{G_N}^{(k)}(\tau_2) +e^{-2\pi i k \bar\tau}\, \C_{G_N}^{(-k)}(\tau_2) \right) \,.
\label{cinst}
\eea
  Since the integrated correlator is real it follows that  $\C_{G_N}^{(k)}(\tau_2) = \C_{G_N}^{(-k)}(\tau_2)$ so that $\C_{G_N}(\tau,\bar\tau)$   contains equal contributions from instantons and anti-instantons.

\subsection{Perturbative contribution}
\label{pertexpn}

The perturbative sectors of the integrated correlators, $\C_{G_N}^{pert}$ derived from
the localised partition function, were discussed  in \cite{Alday:2021vfb},  where they were expressed in terms of generalised  Laguerre polynomials as  reviewed in appendix~\ref{pertcont}. One of the primary interests in \cite{Alday:2021vfb} was to use this perturbative data to determine terms in the large-$N$ expansion order by order in $1/N$ {or, more precisely, order by order in the inverse central charges, }$1/c_{G_N}$.  However,  here we will study the perturbation expressions at finite $N$ in more detail, which will motivate the form of a set of Laplace-difference equations that generalises the analysis in \cite{Dorigoni:2021bvj,Dorigoni:2021guq}  of the $SU(N)$ correlators, as well as the modular covariant expressions \eqref{mainres}  that are well-defined for all values of $N$ and $\tau$. Further strong  evidence for these expressions will be obtained from the evaluation of the instanton contributions  in the next subsection.

Our starting point is the explicit result for the perturbative sector $\C_{G_N}^{pert}$ obtained in \cite{Alday:2021vfb}, and for convenience summarised in appendix \ref{pertlocal}.  The expansions of the expressions in \eqref{pertSUn}-\eqref{pertUSpN}  in powers of $g_{_{YM}}^2$ can be organised in a striking manner by defining the expansion parameters, $a_{G_N}$,  for each gauge group in the following manner\footnote{ Note that the  definition of $a_{G_N}$ differs from that in \cite{Alday:2021vfb}.}
\bea  
a_{SU(N)} = {N g_{_{YM}}^2 \over 4\pi^2}\,, \qquad
a_{SO(n)} ={(n-2)  g_{_{YM}}^2 \over 4\pi^2} \,, \qquad
a_{USp(n)} = {(n+2) g_{_{YM}}^2  \over 8\pi^2} \, ,
\label{pertgroup}
\eea
where $n=2N$ or $2N+1$ for $SO(n)$, and $n=2N$ for $USp(n)$.\footnote{The symbol $n$ is introduced is to unify the formulae for $SO(2N)$ and $SO(2N+1)$, and to show the connection between $USp(2N)$ and $SO(2N)$ correlators. }  We note that $a_{SU(N)}$ is the 't Hooft coupling of the $SU(N)$ theory (up to a factor $4\pi^2$), while $a_{SO(n)}$ and $a_{USp(n)}$ are the generalisations for $SO(n)$ and $USp(n)$ theory (see also \cite{Cvitanovic:2008zz}). \footnote{In the case of $SO(3)$, one needs to rescale $g_{_{YM}} \to \sqrt{2}\, g_{_{YM}}$ and define $a_{SO(3)} = {g_{_{YM}}^2 / (2\pi^2)}$ so that $a_{SO(3)} = a_{SU(2)} = a_{USp(2)}$. See also discussion below \eqref{pertUSpN}. Furthermore, one can see that $a_{SU(4)}=a_{SO(6)}$\ and $a_{USp(4)}=a_{SO(5)}$, consistent with the isomorphic relations among these groups.  }

The perturbative 't Hooft couplings defined in \eqref{pertgroup} can be rewritten in the compact form  $a_{G} = h^\lor_G g_{_{YM}}^2/(4\pi^2)$, with $ {h^\lor_G} $ the dual Coxeter number for the group $G$. The appearance of the dual Coxeter number is quite natural in $\mathcal{N}=4$ SYM when all the fields belong to the adjoint representation.

In terms of these parameters we find that the perturbative expansion of all the integrated correlators can be expressed in the following form, 
\ie
\C_{G_N}^{pert} (\tau_2) =  & \, -4 c_{G_N} \left[ \frac{3   \, \zeta (3) a_{G_N}   }{2} -\frac{75 \, \zeta (5)a_{G_N}^2}{8} 
+\frac{735 \,\zeta (7) a_{G_N}^3}{16} -\frac{6615  \,\zeta (9)  \left(1 + P_{G_N, 1}\right)  a_{G_N}^4 }  {32} \right. \\
& \left. +\, \frac{114345 \,  \zeta (11) \left(1+  P_{G_N, 2}  \right)a_{G_N}^5  }{128 }
 -\frac{3864861 \,\zeta(13) \left(1 +  P_{G_N, 3}  \right) a_{G_N}^6}{1024} \right.  \\ 
& \left. +\,   \frac{32207175 \,\zeta(15) \left(1+ P_{G_N, 4}  \right) \, a_{G_N}^7 }{2048 }+ \mathcal{O}(a_{G_N}^{8}) \right] \, ,
\label{pertexp}
\fe     
where $c_{G_N}$ is the conformal anomaly or central charge associated with $G_N$ and is given by 
\bea
c_{SU(N)}= \frac{N^2-1}{4}\,,\qquad c_{SO(n)}= \frac{n(n-1)} {8}\,, \qquad c_{USp(n)} = \frac{n(n+1)}{8}\, .
\label{confanom}
\eea
We see that the first three perturbative contributions are universal and their dependence on $N$ is contained entirely within $c_{G_N}$ and $a_{G_N}$. Explicit ``non-planar''  factors, $P_{G_N,i}$, where $i = \ell-3$ and $\ell$ is the loop number, first  enter at four loops and the first few examples are listed below:
\begin{itemize} 
\item $SU(N)$
\ie
\label{suparam}
\qquad P_{SU(N), 1}  &= \frac{2}{7N^2} \, , \qquad\qquad P_{SU(N), 2} =  {1 \over N^{2}} \, , \cr
\qquad P_{SU(N), 3} &=  \frac{25N^2 +4}{11 N^4}   \, , \qquad  P_{SU(N), 4} = \frac{605 N^2+332}{143 N^4} \, .
\fe
\item $SO(n)$
\ie
\label{soparam}
P_{SO(n), 1}  &= -\frac{n^2-14 n+32}{14 (n-2)^3}\, , \qquad\qquad P_{SO(n), 2} = -\frac{n^2-14 n+32}{8 (n-2)^3}\, \,, \cr
 P_{SO(n), 3} &=- \frac{12 n^4-221n^3+1158 n^2-2432 n +1856}  {22 (n-2)^5} \, , \cr
   P_{SO(n), 4} &= - \frac{2 \left(342 n^5- 7217 n^4- 48841 n^3 - 153938 n^2   +  239232 n  - 149920\right)}{715 (n-2)^6} \, .
\fe
\item $USp(n)$
\ie
\label{uspparam}
P_{USp(n), 1}  &= \frac{n^2+14 n+32}{14 (n+2)^3}\ \, , \qquad P_{USp(n), 2} = \frac{n^2+14 n+32}{8 (n+2)^3} \, , \cr
 P_{USp(n), 3} &= \frac{12 n^4 + 221n^3+1158 n^2 +2432 n +1856}  {22 (n+2)^5}\, , \cr
   P_{USp(n), 4} &= \frac{2 \left(342 n^5 +7217 n^4+ 48841 n^3 + 153938 n^2   +  239232 n  + 149920\right)}{715 (n+2)^6}  \, . 
\fe
\end{itemize}

Some interesting features of these expansions are as follows.
\begin{itemize}

\item
Whereas the genus expansion of  $SU(N)$ gauge theory in powers of  $1/N^2$ and $a_{SU(N)}$  \cite{tHooft:1973alw}   is well known, there seems to be no systematic analysis in the literature of the analogous  expansions  for $SO(n)$  and $USp(n)$ (although there are some limited results in \cite{Cvitanovic:2008zz}). We see from  \eqref{pertgroup}, \eqref{pertexp}, \eqref{soparam} and \eqref{uspparam}  that these expansions are  purely in powers of 
$1/(n-2)$ and $1/(n+2)$, respectively.   Indeed, if we define the parameters 
\ie\label{Ngenus}
 N_{SU(N)} = N^2 \, , \qquad  N_{SO(n)} = n-2  \, , \qquad  N_{USp(n)} = n+2 \, ,
 \fe
 the expansion \eqref{soparam} can be re-expressed in a form that generalises the topological expansion of the $SU(N)$ case, in which it takes the general form
\bea
\C_{G_N} (\tau, \bar \tau) \sim  \C_{G_N}^{pert}(\tau_2) \sim c_{G_N} \sum_{g=0}^{\infty} (N_{G_N})^{-g} \,  \C^{(g)}_{G_N} (a_{G_N}) \, , 
\label{topo1early}
\eea
where the coefficients\footnote{The seemingly strange choice for $N_{SU(N)}=N^2$ is such that for the case of $SU(N)$ we obtain exactly the standard genus expansion of the form $c_{SU(N)}\sum_{g\geq 0 }N^{-2g} \C^{(g)}_{SU(N)}(a_{SU(N)})$. }  $\C^{(g)}_{G_N} (a_{G_N})$ are power series, with rational coefficients, in the expansion parameter $a_{G_N}$ defined in \eqref{pertgroup}. Following the terminology in the $SU(N)$  case, we will refer to terms with $g\ge 1$ as ``non-planar'' terms.

\item
 A striking property of \eqref{pertexp}  is that  the expression for the planar contribution $\C^{(0)}_{G_N}(a_{G_N})$ is the same for all the groups, and the non-planar contributions only enter at $\ell\geq4$ loops, i.e. $\C^{(1)}_{G_N}(a_{G_N}) = O(a_{G_N}^4)$. Such a property can be seen directly from the construction of perturbative loop integrands using the methods in \cite{Eden:2011we, Eden:2012tu}, and will have important consequences when we consider the large-$N$ expansions. This property is only manifest with definition of the expansion parameters  given in \eqref{pertgroup}.
 
Furthermore, the precise coefficients at each order of the perturbative expansion given in \eqref{pertexp} can be verified using standard quantum field theory results. This calculation was described for the first two loops in \cite{Dorigoni:2021guq} and for the planar terms up to order $O(a_{G_N}^4)$ in \cite{Wen:2022oky} by explicitly performing the relevant higher-loop integrals. These results make use of the perturbative loop integrands constructed in \cite{Eden:2011we, Eden:2012tu, Bourjaily:2015bpz, Bourjaily:2016evz} and the precise expression for the integrated correlator \eqref{intcorr} (see e.g. (2.3) of \cite{Dorigoni:2021guq}).
 
\item We should stress that the definition of the expansion parameters, $N_{G_N}= {h^\lor_{G_N}} g_{_{YM}}^2/(4\pi^2)$ defined in  \eqref{Ngenus}, differ from the parameters that enter in the large-$N$ expansion in the holographic limit, which  will be considered in section~\ref{sec:strongthooft}. In that case the parameters, which  are denoted  $\tilde N_{G_N}$,  are defined in  \eqref{eq:lamval} in terms of the Ramond--Ramond  five-form flux of an orientifold  background. This is reviewed in appendix \ref{sec:stringy}.  It is only if we use the expansion parameters defined in  \eqref{Ngenus} that the weak-coupling perturbative expansion \eqref{pertexp} has a finite number of non-planar terms, i.e. terms that are suppressed by powers of  $1/N_{G_N}$,  at fixed loop order $O(a_{G_N}^\ell)$.

\item
The symmetry under the interchange $(N, g_{YM}^2)\leftrightarrow (-N,-g_{_{YM}}^2$) is evident from the form of \eqref{pertexp}.  For $SU(N)$, we have 
\ie
c_{SU(N)} = c_{SU(-N)} \, , \qquad a_{SU(N)}  = a_{SU(-N)}\, , \qquad  P_{SU(N), i} = P_{SU(-N), i} \, ,
\fe
hence
\ie
\C_{SU(N)}^{pert}(g^2_{_{YM}}) =\C_{SU(-N)}^{pert}(-g^2_{_{YM}})\,,
\fe  
which reflect a relation between $SU(N)$ and $SU(-N)$.  There are also relations between $SO(2N)$ and $USp(-2N)$ under $(N, g_{YM}^2)\leftrightarrow (-N,-2g_{_{YM}}^2$)
\ie
  c_{SO(2N)} = c_{USp(-2N)} \, , \qquad a_{SO(2N)}  = 2 a_{USp(-2N)} \,, \qquad P_{SO(2N), i} = P_{USp(-2N), i} \,,   
\fe 
which lead to 
\ie
\C_{SO(2N)}^{pert}( g^2_{_{YM}}) =\C_{USp(-2N)}^{pert}(-2 g^2_{_{YM}}) \, .
\fe  
These relations have been further checked at higher orders.   We  will  return to this point later in the discussion of the  Laplace-difference equations.

\end{itemize}

\subsection{Yang--Mills instanton sectors}
\label{sec:yminst}

In order to discuss the instanton contributions to $\C_{G_N}$ we will make use of the expressions shown in appendix~\ref{instcont} for the contribution of instantons to the $\mathcal{N}=2^*$ SYM partition function, $\hat Z^{inst}(m,a_i)$ that were obtained in \cite{Billo:2015pjb, Billo:2015jyt}.  In particular, the full non-perturbative sector, presented in \eqref{cinst}, can be computed from
\begin{equation}
\C_{G_N}^{inst} (\tau,\bar\tau) =\tau_2^2\partial_\tau\partial_{\bar\tau} \partial_m^2  Z_{G_N}^{inst}(m, \tau,\bar\tau) \big|_{m\to 0}\,,
\end{equation}
with $ Z_{G_N}^{inst}(m, \tau,\bar\tau)$ the non-perturbative contribution to the localised $\mathcal{N}=2^*$ partition function.
As briefly reviewed in appendix \ref{pertlocal}, $ Z_{G_N}^{inst}$ can be obtained by a suitable matrix model integral over the variables $a_i$ of the Nekrasov  partition function $ \hat Z_{G_N}^{inst} (m, \tau, a_{i}) $.
The $k$-instanton contribution to the Nekrasov  partition function follows from the Fourier sum \eqref{instsum}
 \ie
 \hat Z_{G_N}^{inst} (m, \tau, a_{i}) = \sum_{k=1}^{\infty} e^{2\pi i k \tau}  \hat Z^{(k)}_{G_N}(m, a_{i})+ c.c. \, ,
 \fe   
 where the complex conjugate, indicated by $c.c.$, contains the anti-instanton contribution.

The small-$m$ expansion of the $k$-instanton contribution for $SU(N)$ was well studied in \cite{Chester:2019jas} and led to the following compact expression, 
 \bea
\label{kinstsuN}
 \partial_{m}^2   \hat Z_{SU(N)}^{(k)}(m, a_{i})  \big{|}_{m=0} &= & \sum_{\underset{ p q =k}  {p,q>0}} \oint  {dz \over 2\pi}
\prod^p_{a=1} \prod^q_{b=1} \prod_{j=1}^N  {(z-a_j + i \,k_{a,b} )^2\over (z-a_j + i \,k_{a,b} )^2+1 } \times
\left[
\left({2\over p^2}+{2\over q^2} \right) \right. 
 \\
&& \left. + \sum_{j=1}^N
{i \, (q+p)(q-p)^2 \over pq [z-a_j+i \,(p+q -1)] [z-a_j+ i \,(q-1) ] [z-a_j+i \,(p-1)]}
\right] \,, \nn
\eea
where the integration contour $z$ is a counter-clockwise contour surrounding the poles at $z= a_j + i$
(with $j=1,\dots,N$) and $k_{a,b}= a+b-2 $.     

In appendix~\ref{instcont} we briefly summarise the results of \cite{Billo:2015pjb, Billo:2015jyt} regarding the computation of the instantonic sectors via equivariant supersymmetric localisation for $\cN=4$ SYM with gauge groups  $SO(2N)$, $SO(2N+1)$ and $USp(2N)$.
 Here we only present the results in the special case of  relevance to us,  in which the omega deformation parameters are set to $\epsilon_{1}=\epsilon_{2}=1$, which amounts to localisation on $S^4$.
We will only consider the complete expression in the  single-instanton case ($k=1$), and determine multiple-instanton contributions only for certain particular values of $N$. The general procedure is presented in  appendix~\ref{instcont}, based on  
\cite{Billo:2015pjb, Billo:2015jyt}.  Here we will determine the explicit  small-$m$ expansion of these results, which are relevant for the computation of the integrated correlators.

\begin{itemize}
\item $SO(2N)$:

The one-instanton contribution for $SO(2N)$ is obtained by performing a one-dimensional contour integral using \eqref{eq:othergroup} and \eqref{eq:SO2nInt}. The relevant poles are at  $\phi_1=a_j+\epsilon_+/2, -a_j+\epsilon_+/2, \epsilon_3/2, \epsilon_4/2$ \cite{Billo:2015pjb}. Collecting all these residues, setting $\epsilon_{1}=\epsilon_{2}=1$, and taking small-$m$ expansion, we find,
 \ie
\label{kinstso2N}
\partial_m^2 \hat Z_{SO(2N)}^{(1)}(m, a_{i}) \big{|}_{m=0} = \sum_{j=1}^N \left( R_{a_j+\epsilon_+/2} + R_{-a_j+\epsilon_+/2} \right) + R_{\epsilon_3/2}+ R_{\epsilon_4/2} \, ,
 \fe
 where $R_X$ is the result of taking residue at the pole at $\phi_1 = X$, and they are given by
 \ie
 R_{ \pm a_j+\epsilon_+/2}   &= \frac{ 2( \pm i a_j+1) ( \pm a_j+2)}{( \pm 2 i a_j +3)^2} \prod_{\ell \neq j} \frac{[(\pm i a_j +1)^2 + a^2_{\ell}]^2}{[a^2_{\ell}-a^2_j] [(\pm i a_j +2)^2 + a^2_{\ell}] } \, , \cr
R_{\epsilon_3/2} +R_{\epsilon_4/2}  &= - \partial_m^2 \Big[ \frac{m(m-3)}{32}  \prod_{j=1}^{N} \frac{ 4 a^2_j + (3 m-1)^2  }{ 4 a^2_j + (m-3)^2 } + (m \rightarrow -m) \Big] \Big{|}_{m=0} \, . 
 \fe
In the final expression we have used the continuation $a_j \to i \, a_j$, which will also  be used for the $SO(2N+1)$ and $USp(2N)$ cases considered below.
 
Although we have not considered the general $k$-instanton expression for general $N$, we have evaluated special examples using the prescription {for} contour integrals that {is} discussed in appendix~\ref{instcont}.  For example, the $k=2$ contribution to the integrated correlator in the $SO(4)$  case  has the form
\ie
\partial_m^2 \hat Z_{SO(4)}^{(2)}(m, a_{i})  \big{|}_{m=0} = \frac{51}{16}-\frac{6}{  \left(a^+_{12}\right){}^2+9}-\frac{6}{ a_{12}^2+9}+\frac{12}{\left[\left(a^+_{12}\right){}^2+9\right]{}^2}+ \frac{12}{\left[a_{12}^2+9\right]{}^2} \, ,
   \fe
   where $a_{ij} = a_i -a_j$ and $a_{ij}^+ = a_i+a_j$.

   \item $SO(2N+1)$: 
   
   The computation for  $SO(2N+1)$ is similar.  In this case the one-instanton contribution for general $N$ is given by 
 \ie
\label{kinstso2N1}
\partial_m^2 \hat Z_{SO(2N+1)}^{(1)}(m, a_{i}) \big{|}_{m=0} = \sum_{j=1}^N \left( R_{a_j+\epsilon_+/2} + R_{-a_j+\epsilon_+/2} \right) + R_{\epsilon_3/2}+ R_{\epsilon_4/2} \, .
 \fe
Again $R_X$ is the result of taking residue at the pole at $\phi_1 = X$,  and each takes the following form,  
 \ie
  R_{ \pm a_j+\epsilon_+/2}   &= \frac{ 2(\pm i a_j+1)^3}{ \pm  i a_j (\pm 2 i a_j+3)^2}  \prod_{\ell \neq j}  \frac{\left[ (\pm i a_j+1)^2+a_{\ell}^2\right]^2}{\left(a_{\ell}^2-a_{j}^2\right) \left[ (\pm i a_j+2)^2+a_{\ell}^2\right]}\, , \cr
R_{\epsilon_3/2} +R_{\epsilon_4/2}  &=   - \partial_m^2 \Big[ \frac{m(3m-1)}{32}  \prod_{j=1}^{N} \frac{ 4 a^2_j + (3 m-1)^2  }{ 4 a^2_j + (m-3)^2 } + (m \rightarrow -m) \Big] \Big{|}_{m=0} \, .
 \fe

As a special example we have evaluated the $k=2$ contribution for the  $SO(5)$ theory, which has the form
\ie
\partial_m^2 \hat Z_{SO(5)}^{(2)}(m, a_{i})  \big{|}_{m=0} &=\Bigg( \frac{47}{32}-\frac{2}{a_1^2+4}+\frac{4}{\left(a_1^2+4\right)
   \left(a_2^2+4\right)}\cr
& ~~~~  -\frac{6 \left[2 \left(a_1^4-\left(a_2^2-21\right)
   a_1^2+7 a_2^2+151\right)
   a_1^2+387\right]}{\left[a_1^4-2
   \left(a_2^2-9\right)
   a_1^2+\left(a_2^2+9\right)^2\right]^2} \Bigg) + (a_1 \leftrightarrow a_2)\,.
   \fe

 \item $USp(2N)$:  
  
For the  $USp(2N)$ group, the number of contour integrals is equal to $\lfloor \frac{k}{2}\rfloor$, with $k$ the instanton number. Therefore,  no contour integral is involved in the one-instanton case. For this reason, the one-instanton contribution {is given by} the following compact expression, 
 \ie
\label{kinstsp2N}
\partial_m^2 \hat Z_{USp(2N)}^{(1)}(m, a_{i}) \big{|}_{m=0} = {1\over 2}  \prod_{j=1}^N \frac{a_j^2}{a_j^2+2} \, .
 \fe

When $k=2$ and $k=3$, the contour integrals are only one-dimensional, and are relatively easy to perform. For instance, the two- and three-instanton contributions for $USp(4)$ are found to be: 
\ie
\partial_m^2 \hat Z_{USp(4)}^{(2)}(m, a_{i})\big{|}_{m=0}  &=\Bigg(\frac{19}{16} -\frac{6}{2
   a_1^2+9}+\frac{12}{\left[2
   a_1^2+9\right]^2}-\frac{96}{\left(2
   a_1^2+9\right) \left[2
   a_2^2+9\right]^2} \cr
   &~~~~ -\frac{8 \left[4 \left(a_2^2+3\right) a_1^4+
   \left(32 a_2^2+57\right) a_1^2-30\right]}{\left(2
   a_1^2+9\right)    \left(2 a_2^2+9\right) \left(a_{12}^2+8\right)
   \left[(a_{12}^+)^2+8\right]} \Bigg)+ (a_1 \leftrightarrow a_2) \, , 
   \fe
   and      
      \ie
\partial_m^2 \hat Z_{USp(4)}^{(3)}(m, a_{i}) \big{|}_{m=0}  &= \frac{ \left(a_2^4+20 a_2^2+80\right) a_1^4+100 \left(a_2^2+8\right) a_1^2+1024}{3
   \left[a_1^2+8\right]^2 \left[a_2^2+8\right]^2} + (a_1 \leftrightarrow a_2)\,.
   \fe
We have also computed $\partial_m^2 \hat Z_{USp(2N)}^{(k)}(m, a_{i})\big{|}_{m=0} $ for $USp(2N)$ for $k=2,3$, with $2\le N\le5$. However, some of the expressions are somewhat lengthy and we will not show them explicitly here. 

\end{itemize}

Once $\partial_m^2 \hat Z_{G_N}^{(k)}(m, a_{i})\big{|}_{m=0}$ has been determined, it is straightforward to compute the matrix integrals using the expressions for  expectation values given in \eqref{SO1exp}, \eqref{SO2exp} and \eqref{USpexp}.  We find the resulting instanton contributions to $\C^{inst}_{G_N}(\tau, \bar \tau)$ agree precisely with the expected results based on the duality-covariant ansatz  \eqref{mainres}.  We will return to this comparison in section~\ref{sec:ansatz} where we will discuss the ansatz and its motivation in more detail.

\section{Laplace-difference equations}
\label{sec:lapdiff}

A striking property of the formulation of the $SU(N)$ integrated correlator in  \cite{Dorigoni:2021bvj,Dorigoni:2021guq} is that it satisfies a Laplace equation that relates it to the $SU(N-1)$ and $SU(N+1)$ correlators,
\begin{align}
\Delta_\tau \C_{SU(N)}(\tau,\bar{\tau})   -4c_{SU(N)} & \Big[\C_{SU(N+1)}(\tau,\bar{\tau})- 2\,\C_{SU(N)}(\tau,\bar{\tau})+\C_{SU(N-1)}(\tau,\bar{\tau})\Big]\nn\\
  &-(N+1)\,\C_{SU(N-1)}(\tau,\bar{\tau}) + (N-1)\, \C_{SU(N+1)}(\tau,\bar{\tau})=0 \, .
\label{lapdiffSUN}
\end{align}
This equation, which is reviewed in appendix~\ref{sec:laplace-dif}, has powerful consequences. Given the initial condition $\C_{SU(1)}=0$, this equation easily determines the correlator for gauge group $SU(N)$ in terms of the correlator for  gauge group $SU(2)$.  Furthermore it gives a very simple iterative procedure for determining terms in the large-$N$ expansion of the  correlator for gauge group $SU(N)$.  We will now see how these statements generalise to any of the classical Lie groups.

Our procedure is to determine the Laplace-difference equations for general classical gauge groups  by requiring consistency with the expressions determined in the previous section supplemented with the requirement of consistency with GNO duality.  Using the perturbative results given in the section  \ref{intgencorr}, we find that the integrated correlators obey equations of the form \eqref{lapdiffGN},   in which the coefficients $d_{G_{N-1}}$, $d_{G_{N}}$ and  $d_{G_{N+1}}$ are determined.  Explicitly, we find the Laplace-difference equation for $SO(n)$ (with $n=2N$ or $n=2N+1$) is given by (more discussion of these equations  is given in  appendix~\ref{sec:laplace-dif})
\begin{align}
\Delta_\tau \C_{SO(n)}(\tau, \bar\tau)  -  2 c_{SO(n)} & \Big[ \C_{SO(n+2)}(\tau, \bar\tau) -2 \,\C_{SO(n)}(\tau, \bar\tau) +\C_{SO(n-2)} (\tau, \bar\tau)  \Big] \nn\\
& - n\, \C_{SU(n-1)} (\tau, \bar\tau) +(n-1)\, \C_{SU(n)} (\tau, \bar\tau) =0 \, .
\label{lapdiffSO}
\end{align} 
The Laplace-difference equation for $USp(n)$ (with $n=2N$) takes a very similar form, 
\begin{align}
\Delta_\tau \C_{USp(n)}(\tau,\bar\tau) - 2 c_{USp(n)} & \Big[ \C_{USp(n+2)}(\tau,\bar\tau) -2 \,\C_{USp(n)}(\tau,\bar\tau)+\C_{USp(n-2)} (\tau,\bar\tau) \Big]\nn\\
&  +   n\, \C_{SU(n+1)} (2\tau,2\bar\tau)- (n+1)\,\C_{SU(n)} (2\tau,2\bar\tau)   =0\, . 
\label{lapdiffUSp}
\end{align}
Note that there is an important rescaling $(\tau, \bar\tau) \to (2\tau,2\bar\tau)$  in the $SU(N)$ correlators in the second line of \eqref{lapdiffUSp}.    
\\

{\bf Lemma}. {\it Equations \eqref{lapdiffSUN} - \eqref{lapdiffUSp} can be solved iteratively to determine $\C_{G_N}$  for any classical Lie group $G_N$, once $\C_{SU(2)}(\tau,\bar\tau)$ is given. }\\

{\it Proof}.  The proof follows from identities satisfied by $\C_{G_N}(\tau,\bar\tau)$ for small values of $N$. 
\begin{itemize}

\item
As discussed in \cite{Dorigoni:2021bvj,Dorigoni:2021guq}, the fact that the  integrated correlator $\C_{SU(1)}=0$ implies that the equation for $\C_{SU(N)}$ \eqref{lapdiffSUN} can be solved for any $N$ in terms of $\C_{SU(2)}$ .

\item
{ The solutions   for other groups follow by use of the identities:
$\C_{USp(0)}= \C_{SO(0)} =\C_{SO(1)}= \C_{SO(2)} =0$.  Equation \eqref{lapdiffSO} with $n=2$ and the fact that $\C_{SO(2)}=0$ determine $\C_{SO(4)}$. Using $n=2$ in \eqref{lapdiffUSp} and $\C_{USp(2)}(\tau,\bar{\tau})=\C_{SU(2)}(\tau,\bar{\tau})$  determines $\C_{USp(4)}$.  Similarly,  \eqref{lapdiffSO}  with $n=3$ and the fact that $\C_{SO(3)}(\tau,\bar{\tau})=\C_{SU(2)}(2\tau,2\bar{\tau})$ (remembering that the localised $SO(3)$ correlator is actually $\C_{SO(3)}(\frac{\tau}{2},\frac{\bar{\tau}}{2})$ as discussed earlier) determine $\C_{SO(5)}$,}\footnote{It should be emphasised that the initial conditions $\C_{SU(2)}(\tau,\bar{\tau})=\C_{SO(3)}(2\tau,2\bar{\tau}) = \C_{USp(2)}(\tau,\bar{\tau}) $ are non-trivial properties.
Using \eqref{pertSUn}, \eqref{pertsoodd}, and \eqref{pertUSpN},  it is easy to check that their perturbative components are identical, and we have also confirmed that their non-perturbative terms agree.}

\item
Given the above initial conditions for small values of $N$, the solutions for arbitrary $N$ follow iteratively from the equations.

\end{itemize}

We can now consider a few examples of the solutions to the Laplace-difference equations using the procedure outlined above. This will help us to better understand the structures of the correlators and motivates a general ansatz for the integrated correlators, which we will discuss more detail in the next section. The general expression of $\C_{SU(N)}$, which may be obtained from \eqref{lapdiffSUN}, was given in \cite{Dorigoni:2021bvj,Dorigoni:2021guq}. Here we will consider the correlators in other gauge groups and use the general result of $\C_{SU(N)}$. 

Let us begin with the correlators for $SO(2N)$.  Using \eqref{lapdiffSO}, it is straightforward to  show that
\ie
\C_{SO(4)}(\tau,\bar\tau) &= 2\, \C_{SU(2)}(\tau,\bar\tau) \, , \qquad \C_{SO(6)}(\tau,\bar\tau) =  \C_{SU(4)}(\tau,\bar\tau) \, , \cr
\C_{SO(8)}(\tau,\bar\tau) &= -2\, \C_{SU(2)}(\tau,\bar\tau )+\frac{8}{3} \C_{SU(3)}(\tau,\bar\tau )-2 \,\C_{SU(4)}(\tau,\bar\tau )+\frac{4}{5} \C_{SU(5)}(\tau,\bar\tau)+\frac{2}{3} \C_{SU(6)}(\tau,\bar\tau)\, ,
\fe
where we have used $\C_{SO(2)}(\tau,\bar\tau) =0$.  The expressions for $\C_{SO(4)}(\tau,\bar\tau)$ and $\C_{SO(6)}(\tau,\bar\tau)$ reflect the relations $SO(4) \cong SU(2) \times SU(2)$ and $SO(6)  \cong SU(4)$, respectively. It is easy to see from the structure of the Laplace-difference equation that $\C_{SO(2N)}(\tau,\bar\tau)$ can be expressed in terms of linear combination of $\C_{SU(m)}(\tau,\bar\tau)$ with $m=2, 3, \ldots, 2N-2$, as in the example of $\C_{SO(8)}(\tau,\bar\tau)$ given above. As shown in \cite{Dorigoni:2021bvj,Dorigoni:2021guq},  $\C_{SU(m)}(\tau,\bar\tau)$ may be expressed as an infinite sum of the non-holomorphic Eisenstein series $E(s; \tau,\bar\tau)$, or equivalently a two-dimensional lattice sum, hence the same is also true for $\C_{SO(2N)}(\tau,\bar\tau)$, which we will discuss in more detail in the next section. 

We now consider the  integrated correlators in the $SO(2N+1)$ and $USp(2N)$ cases.  We will see that the expressions for these correlators are related by  GNO duality. To begin we will  consider the first non-trivial correlators, $\C_{SO(5)}$ and $\C_{USp(4)}$. The Laplace-difference equations allow us to express the correlators in terms of the $SU(N)$ correlators, 
\ie
\C_{SO(5)}(\tau,\bar\tau)= \left[ -2\, \C_{SU(2)}(\tau,\bar\tau )  + \frac{4}{3}\C_{SU(3)}(\tau,\bar\tau ) \right] +  \left[-2\, \C_{SU(2)}(2 \tau,2\bar\tau
   ) +\frac{4}{3} \C_{SU(3)}(2
   \tau,2\bar\tau ) \right] \, ,
   \label{so5cor}
\fe
with an identical result for $\C_{USp(4)}(\tau,\bar\tau)$, reflecting the fact that $USp(4) \cong SO(5)$. Using the results for $\C_{SU(N)}(\tau,\bar\tau)$, we find $\C_{SO(5)}(\tau,\bar\tau)$ (or equivalently  $\C_{USp(4)}(\tau,\bar\tau)$) can be also be expressed in terms of infinite sums of non-holomorphic Eisenstein series, but importantly involving both $E(s; \tau,\bar\tau)$ and $E(s; 2\tau,2\bar\tau)$. 

We will now consider $\C_{SO(7)}$ and $\C_{USp(6)}$, which will suggest  the general structure of the integrated correlators and the GNO duality that relates $\C_{SO(2N+1)}$ and $\C_{USp(2N)}$. From  \eqref{lapdiffSO} we find that  $\C_{SO(7)}$ is given as a sum of $\C_{SU(N)}$ correlators of the form
\ie
\C_{SO(7)}(\tau,\bar\tau) &= \left[\frac{8}{5} \C_{SU(2)}(\tau,\bar\tau )-\frac{12}{5} \C_{SU(3)}(\tau,\bar\tau
   )+\frac{3}{5} \C_{SU(4)}(\tau,\bar\tau )+\frac{4}{5} \C_{SU(5)}(\tau,\bar\tau ) \right] \cr
   & ~~ + \left[ \frac{3}{5}\C_{SU(2)}(2 \tau,2\bar\tau )-\frac{12}{5} \C_{SU(3)}(2 \tau,2\bar\tau )+\frac{8}{5} \C_{SU(4)}(2 \tau,2\bar\tau) \right]\, ,
   \label{so7cor}
\fe
and from \eqref{lapdiffUSp} $\C_{USp(6)}$ we have
\ie
\C_{USp(6)}(\tau,\bar\tau) &= \left[\frac{8}{5} \C_{SU(2)}(2\tau,2 \bar\tau )-\frac{12}{5} \C_{SU(3)}(2\tau,2\bar\tau
   )+\frac{3}{5} \C_{SU(4)}(2\tau, 2\bar\tau )+\frac{4}{5} \C_{SU(5)}(2\tau,2\bar\tau ) \right] \cr
   &~~  + \left[ \frac{3}{5}\C_{SU(2)}(\tau,\bar\tau )-\frac{12}{5} \C_{SU(3)}(\tau,\bar\tau )+\frac{8}{5} \C_{SU(4)}(\tau,\bar\tau) \right]\, .
   \label{USp6cor}
\fe
Since $ \C_{SU(N)}(\tau,\bar\tau)= \C_{SU(N)}(-\frac{1}{\tau},-\frac{1}{\bar\tau})$ and $\C_{SU(N)}(2\tau, 2\bar\tau) =\C_{SU(N)}(-\frac{1}{2\tau},-\frac{1}{2\bar\tau})$, it follows from \eqref{so7cor} and \eqref{USp6cor} that under the transformation $\hat S: \tau \to   - 1/(2\tau) $, $\C_{SO(7)}(\tau,\bar\tau)$ transforms into  $\C_{USp(6)}(\tau,\bar\tau)$. More generally, by induction, using the Laplace-difference equations \eqref{lapdiffSO} and \eqref{lapdiffUSp},  one can prove 
\ie \label{eq:GNO}
\C_{SO(2N+1)}(\tau, \bar{\tau}) = \C_{USp(2N)}\Big(-\frac{1}{2\tau},-\frac{1}{2\bar\tau}\Big) \, ,
\fe
which is the statement of GNO duality (recalling our previous comment that for $N=1$ the localised correlator equals $\C_{SO(3)}(\frac{\tau}{2},\frac{\bar{\tau}}{2})$, which also coincides with the modular invariant $\C_{SU(2)}(\tau,\bar{\tau}) = \C_{USp(2)} (\tau,\bar{\tau})$).
This property will be made manifest in the duality covariant ansatz of these correlators that will be proposed in the next section. It is also of note that these Laplace-difference  equations are consistent with the dualities  $\C_{SU(N)}(\tau, \bar\tau  ) = \C_{SU(-N)}(-\tau,-\bar\tau )$ and $\C_{SO(2N)}(\tau,\bar\tau) = \C_{USp(-2N)}(-\frac{\tau}{2}, -\frac{\bar\tau}{2})$, which explicitly hold in perturbation theory, as we discussed earlier. 

%\newpage

\section{The duality covariant ansatz}
\label{sec:ansatz}

In this section we will motivate the conjectured expression for $\C_{G_N}$ as the lattice sum \eqref{mainres}.  
The argument for this expression will be based on the  examples of solutions to the Laplace-difference equations presented in the previous section, which make it clear that the integrated correlators $\C_{SO(2N)}$,  $\C_{SO(2N+1)}$ and $\C_{USp(2N)}$ can be written as linear combinations of $\C_{SU(m)}$ for certain values of  $m$.  The fact that   $\C_{SU(N)}$ can be expressed  (at least formally) as an infinite sum of non-holomorphic Eisenstein series   \cite{Dorigoni:2021bvj,Dorigoni:2021guq}  suggests that $\C_{G_N}$ can also be expressed in terms of  sums of Eisenstein series for any $G_N$.  More precisely,  we will find that  $\C_{SO(2N)}$ is given by an infinite sum of $E(s; \tau, \bar \tau)$, whereas  $\C_{SO(2N+1)}$ and  $\C_{USp(2N)}$ involve both $E(s; \tau, \bar \tau)$ and $E(s; 2\tau, 2\bar \tau)$. 

\subsection{Review of $\C_{SU(N)}$}

In \cite{Dorigoni:2021bvj,Dorigoni:2021guq} it was argued that  the integrated correlator of $SU(N)$ theory can formally  be expressed as an infinite sum of  non-holomorphic Eisenstein  series,
\bea
 \label{eq:SUrev}
\C_{SU(N)}(\tau, \bar \tau) = \frac{N(N-1)}{8}+ \sum_{s=2}^{\infty} b_{SU(N)}(s) E(s; \tau, \bar \tau) \, ,
\eea
where the coefficients $b_{SU(N)}(s)$ are defined in terms of $B_{SU(N)}(t)$ by \eqref{Bexpand} and we have used $ b_{SU(N)}(0) = -N(N-1)/8$.

In our normalisation, a non-holomorphic Eisenstein series is defined by
\begin{align}
\label{eisendef}
E(s; \tau, \bar \tau)  = \sum_{(m,n) \neq (0,0)} {1\over \pi^s} {\tau_2^s \over |m+n \tau|^{2s}}\,,
\end{align}
which has the Fourier series expansion
\begin{align} 
\label{eq:fourierE}
E(s; \tau, \bar \tau) = &\,\,  \frac {2\zeta(2s)}{\pi^s} \tau_2^s  +   \frac{2\sqrt \pi \,\Gamma(s-\half) \zeta(2s-1)}{\pi^s \Gamma(s)}\, \tau_2^{1-s}\\
&\notag+ \sum_{    \underset  {k\ne0} {k=-\infty}    }^\infty e^{2\pi i k\tau_1}    \frac{4 \sqrt{\tau_2}}{\Gamma(s)}\,  |k|^{s-\half}\sigma_{1-2s} (|k|)     \, K_{s-\half}(2\pi |k| \tau_2)\,,
\end{align}
with $K_{s}$ a modified Bessel function of second kind and $\sigma_{s}(k) = \sum_{d\vert k} d^s$ a divisor function.
The expression \eqref{eq:SUrev} is formal since it is not a convergent series, but it can be defined in a convergent manner  by using the integral representation for the Eisenstein series 
\bea
\label{eisendef}
E(s; \tau, \bar \tau) =  \sum_{(m,n) \neq (0,0)}  \int_0^{\infty} e^{-t \pi  {|m+n \tau|^2 \over \tau_2}} {t^{s-1} \over \Gamma(s)} dt \, .
\eea
Substituting in  \eqref{eq:SUrev} and using \eqref{intcon}, gives a well-defined two-dimensional lattice sum expression \cite{Dorigoni:2021bvj, Dorigoni:2021guq},  
\bea
 \label{eq:SUrevB}
\C_{SU(N)}(\tau, \bar \tau) = \sum_{(m,n) \in \Z^2}  \int_0^{\infty} e^{-t \pi  {|m+n \tau|^2 \over \tau_2}}  B_{SU(N)}(t) dt \, , 
\label{CNexp}
\eea
where $B_{SU(N)}(t)$ is a rational function, 
\bea \label{eq:BSUN}
B_{SU(N)}(t) =  \sum_{s=2}^{\infty} b_{SU(N)}(s) {t^{s-1} \over \Gamma(s)} = \frac{\cQ_{SU(N)}(t)}{(t+1)^{2N+1}}\, ,
\eea
and  $\cQ_{SU(N)}(t)$ is a polynomial of degree $(2N-1)$ that  takes the form
 \begin{align}
\cQ_{SU(N)}(t)
&\notag= -{1\over 4} N (N-1) (1-t)^{N-1} (1+t)^{N+1}  \\
 & \left\{ \left(3+  (8N+3t-6) \, t\right ) P_N^{(1,-2)} \left(\frac{1+t^2}{1-t^2}\right)  + \frac{1}  {1+t}   \left(3t^2-8Nt-3 \right) P_N^{(1,-1)}    \left(\frac{1+t^2}{1-t^2}\right)  \right\} \,,
\label{polydef}
\end{align}
with $P_N^{(\alpha,\beta)} (z)$ being a Jacobi polynomial. 
It is notable that $B_{SU(-N)}(t) = B_{SU(N)}(-t)$ which is directly connected to the relation $\C_{SU(N)}(\tau,\bar{\tau}) = \C_{SU(-N)}(-\tau,-\bar{\tau})$.

A key feature of the  function   $B_{SU(N)}(t)$ in the representation of $\C_{SU(N)}$ in  \eqref{eq:SUrevB} is the inversion symmetry $B_{SU(N)}(t) = t^{-1}\, B_{SU(N)}(t^{-1})$.  This property  leads to particular relationships between the coefficients $b_{SU(N)}(s)$ in \eqref{Bexpand} that have important consequences.  In particular, consider the zero Fourier mode  of \eqref{eq:SUrev} (the perturbative sector), $\C_{SU(N)}^{pert}(\tau_2)$, which is the sum of infinitely many zero modes of Eisenstein series.  From \eqref{eq:fourierE}, we see that this results in the sum of two infinite series: 
\bea
\label{eq:cpertdef}
\C_{SU(N)}^{pert}(\tau_2) = \C_{SU(N)}^{(i)}(\tau_2)+ \C_{SU(N)}^{(ii)}(\tau_2)\,  ,
\eea
where $\C_{SU(N)}^{(i)}(\tau_2)$ denotes the sum of $\tau_2^{1-s}$ terms, which is asymptotic and gives a well-defined perturbative series for small $g_{_{YM}}^2 = 4 \pi/\tau_2$, and  $\C_{SU(N)}^{(ii)}(\tau_2)$, which is the sum of $\tau_2^s$ terms, is divergent term by term as $g_{_{YM}}^2\to 0$.  However, as presented in more detail in \cite{Dorigoni:2021guq}, the latter series can be Borel resummed and the result is  
\bea 
\label{eq:pertres1}
\C_{SU(N)}^{(ii)}(\tau_2) = \C_{SU(N)}^{(i)}(\tau_2) =\frac{1}{2} \C_{SU(N)}^{pert}(\tau_2) \,,
\eea
we stress that the lattice sum representation  \eqref{eq:SUrevB} is a well-defined function for all values of $\tau$ in the upper-half plane and for all values of $N\geq0$, while the need for Borel resummation only arises when it is expanded in perturbation theory.

The coefficients $b_{SU(N)}(s)$ in the ansatz  \eqref{eq:SUrev} or, equivalently, the rational function $B_{SU(N)}(t)$, are uniquely determined by matching $\C^{pert}_{SU(N)}(\tau_2)$ with the perturbative terms determined by localisation.
In other words, $B_{SU(N)}(t)$ can be determined by matching  with the perturbative contributions  given in \eqref{pertSUn},  in a manner that we will now describe. 

Let us begin by considering the perturbative contribution arising from the lattice-sum representation. 
Assume that $B_{SU(N)}(t)$ is given by a convergent expansion
\ie
B_{SU(N)}(t) = \sum_{s=1}^\infty \alpha (s)  \, t^s \,.
\fe
Substituting this expression in \eqref{eq:SUrevB} and computing the perturbative terms (in terms of the representation \eqref{eq:SUrev} we are using the fact that the $\tau_2^s$ terms sum to give the same contribution as the $\tau_2^{1-s}$ terms, as discussed above), one finds that the perturbative contribution is given by the asymptotic formal power series
\ie \label{eq:1P}
\C^{pert}_{SU(N)}( y) \sim \sum_{s=1}^\infty \alpha(s) {4 \Gamma\left(s+{1\over 2} \right) \zeta(2s+1) \over \sqrt{\pi}} y^{-s} \, ,
\fe
with $y = \pi \tau_2 = 4 \pi^2 / g_{_{YM}}^2$. We will now compare this result with the perturbative terms obtained from \eqref{pertSUn}.  For convenience we will  denote the perturbative contribution to $\C_{SU(N)}(\tau, \bar \tau)$ in \eqref{pertSUn}  as
\ie \label{eq:3P}
\C^{pert}_{SU(N)}(y) =  \int_0^{\infty}  {d \omega \over \sinh^2 \omega}  \omega\,  y^2 \partial^2_y K(\omega^2/y) \, ,
\fe
and the integrand {has} the following convergent power series expansion
\ie
 \omega\, y^2 \partial^2_y K(\omega^2/y) =  \sum_{s=1}^\infty \beta(s)\, \omega^{2s+1} y^{-s}\, .
\fe
Using the integral identity
\bea
\label{laplactr}
 \int_0^\infty d\omega\frac{\omega^{m+1}}{\sinh^2\omega} =2^{-m} \Gamma(m+2)\zeta(m+1) \, , 
\eea
valid for $m\geq 1$, we obtain the perturbative contribution from \eqref{eq:3P}, which is given by the asymptotic power series
\ie \label{eq:2P}
\C^{pert}_{SU(N)}( y) \sim \sum_{s=1}^\infty \beta(s) { \Gamma\left(2s+2 \right) \zeta(2s+1) \over 2^{2s}} y^{-s} \, .
\fe
By equating \eqref{eq:1P} and \eqref{eq:2P}, we find the following relation between $\alpha(s) $ and $\beta(s)$ 
\bea \label{eq:relation}
\beta(s) = \frac{4 \,\alpha(s) }{  (2s+1)\Gamma(s+1)}  \, . 
\eea
Therefore knowing $y^2 \partial^2_y K(\omega^2/y)$ allows us to determine $B_{SU(N)}(t)$. Explicitly, from \eqref{eq:relation}, we find the following simple relationship\footnote{Note that this procedure is closely related to the $SL(2,\Z)$ Borel transform introduced in \cite{Collier:2022emf}.}
\ie
B_{SU(N)}(t)  &=   {1\over 4}  \int_0^{\infty} dr \,  e^{-r} \, \partial_\omega  \Big[  \omega\, y^2 \partial^2_y K(\omega^2/y) \Big] \Bigg{|}_{y=1, \omega=\sqrt{r t}} \, .
\fe
Using the expression for $K(\omega^2/y)$ given in \eqref{pertSUn}, and  after a suitable change of variables and integration by parts, the above expression can be recast into the following simpler form,
\ie \label{eq:SUNnew}
B_{SU(N)}(t)  &= - t  \int_0^{\infty} dx \,  e^{-x t} \tilde{B}_{SU(N)}(x)  \,   , 
\fe
where the integrand $ \tilde{B}_{SU(N)}(x)$  is directly related to the perturbative result given in \eqref{pertSUn},
\ie \label{eq:SUNnew2}
\tilde{B}_{SU(N)}(x)  &= \, {x^{{3\over 2}} \over 4} \partial_x \Big\lbrace  x^{{3\over 2}} \partial_{x} \Big[ e^{-{x}} \sum_{i,j=1}^N \left( L_{i-1}\left({ x}\right)L_{j-1}\left({ x}\right)-(-1)^{i-j}L_{i-1}^{j-i}\left({x}\right)L_{j-1}^{i-j}\left({ x}\right)\right) \Big] \Big\rbrace  \, .
\fe
Although proving that \eqref{eq:SUNnew} is equivalent to \eqref{eq:BSUN} for arbitrary $N$ is rather non-trivial, it is straightforward to check explicitly the equivalence of these two expressions for any given $N$. We also note that the above derivation is general and not restricted to the $SU(N)$ case, therefore we will apply the same arguments for other gauge groups in the next subsection.

\subsection{Exact expressions for $\C_{SO(2N)}, \C_{SO(2N+1)}$ and $\C_{USp(2N)}$}
\label{sec:exact-exp}

This discussion generalises to the other classical groups.  As described in the previous section, the study of Laplace-difference equations makes it clear that the integrated correlators $\C_{SO(2N)}$,  $\C_{SO(2N+1)}$ and $\C_{USp(2N)}$ can all be expressed as sums of Eisenstein series,  as in the case of $SU(N)$.  More precisely, the analysis of Laplace-difference equations suggests the following ansatz for the integrated correlator for each gauge group 
\ie
\C_{SO(2N)}(\tau, \bar \tau) =\frac{N(N-1)}{4}+ \sum_{s=2}^{\infty} b_{SO(2N)}(s) E(s; \tau, \bar \tau) \, , 
\fe
and\footnote{In the case of $\C_{SO(3)}$, $b^1_{SO(3)}(s)=0$ and $b^2_{SO(3)}(s) = b_{SU(2)}(s)$, and to retrieve the localised correlator we should rescale $(\tau, \bar \tau)\rightarrow (\frac{\tau}{2}, \frac{\bar \tau}{2})$, so that $\C_{SO(3)}(\frac{\tau}{2}, \frac{\bar \tau}{2}) = \C_{SU(2)}(\tau, \bar \tau)$. } 
\ie
\C_{SO(2N+1)}(\tau, \bar \tau) &=  \frac{N^2}{4}  + \sum_{s=2}^{\infty} \left( b^1_{SO(2N+1)}(s) E(s; \tau, \bar \tau) + b^2_{SO(2N+1)}(s) E(s; 2\tau, 2\bar \tau) \right) \, , \cr
\C_{USp(2N)}(\tau, \bar \tau) &= \frac{N^2}{4}  +  \sum_{s=2}^{\infty} \left( b^1_{USp(2N)}(s) E(s; \tau, \bar \tau) + b^2_{USp(2N)}(s) E(s; 2\tau, 2\bar \tau) \right) \, ,
\fe
and \eqref{eq:GNO} implies 
\ie
b^1_{SO(2N+1)}(s) = b^2_{USp(2N)}(s) \, , \qquad b^2_{SO(2N+1)}(s) = b^1_{USp(2N)}(s) \, ,
\fe
since $\hat S$ exchanges $E(s; \tau, \bar \tau)$ with $E(s; 2\tau, 2\bar \tau)$. 
Similarly to $SU(N)$, for the constant term we have used the results $ b_{SO(2N)}(0) = -N(N-1)/4$ and $b_{SO(2N+1)}(0) = b_{USp(2N)}(0) =- N^2/4$.

As in the case of $\C_{SU(N)}(\tau, \bar \tau)$, these formal expressions are well-defined upon using the lattice sum representation of $E(s;\tau,\bar\tau)$ in \eqref{eisendef}, which, using \eqref{intcon}, leads to
\bea \label{eq:SOeven}
\C_{SO(2N)}(\tau, \bar \tau) = \sum_{(m,n) \in \Z^2}  \int_0^{\infty} e^{-t \pi  {|m+n \tau|^2 \over \tau_2}}  B_{SO(2N)}(t) dt \, , 
\eea
where 
\ie
B_{SO(2N)}(t) = \sum_{s=2}^{\infty} b_{SO(2N)}(s) {t^{s-1} \over \Gamma(s)} \, .
\fe 
Similarly, for $SO(2N+1)$ and $USp(2N)$, again using \eqref{intcon}, we have
\begin{align} \label{eq:SOodd}
\C_{SO(2N+1)}(\tau,\bar\tau) &=   \sum_{(m,n) \in \Z^2}  \int_0^\infty dt\left(B^1_{SO(2N+1)}(t) e^{-t \pi\frac{ |m+n\tau|^2}{\tau_2}}+B_{SO(2N+1)}^2 (t) e^{-t\pi \frac{ |m+ 2 n \tau|^2}{2 \tau_2}}\right)\, , 
\end{align}
and 
\begin{align} \label{eq:USpn}
\C_{USp(2N)}(\tau,\bar\tau) &=    \sum_{(m,n) \in \Z^2}  \int_0^\infty dt\left(B^1_{USp(2N)}(t) e^{-t \pi\frac{ |m+n\tau|^2}{\tau_2}}+B_{USp(2N)}^2 (t) e^{-t\pi \frac{ |m+ 2 n \tau|^2}{2 \tau_2}}\right) \, ,
\end{align}
with $B^1_{SO(2N+1)}(t) = B_{USp(2N)}^2 (t)$ and $B^2_{SO(2N+1)}(t) = B_{USp(2N)}^1 (t)$, reflecting GNO duality.

The coefficients $b^i_{G_N}(s)$, or equivalently the rational functions $B^i_{G_N}(t)$, can again  be determined by directly comparing the ansatz with the perturbative results. They can also be fixed using the Laplace-difference equations together with the expression for $\C_{SU(N)}(\tau,\bar\tau)$. Either way, we find   in the $SO(2N)$ case,
\bea
B_{SO(2N)}(t) = \frac{\cQ_{SO(2N)}(t)}{(t+1)^{4N-3}}\, ,
\label{bsondef}
\eea
where  $\cQ_{SO(2N)}(t)$ is a palindromic polynomial  of degree-$(4N-5)$. The following are specific examples,  
\ie \label{eq:SON-ex}
\cQ_{SO(4)}(t) &=2 \cQ_{SU(2)}(t) =     3\, t  (3t^2 -10t+3) \, , \\
 \cQ_{SO(6)}(t) &=\cQ_{SU(4)}(t) = 15\, t \left(3 t^6-23 t^5+50 t^4-72 t^3+50 t^2-23
   t+3\right)  \, , \cr
   \cQ_{SO(8)}(t) &= 126\,  t \left(t^{10}-12 t^9+47 t^8-122 t^7+167
   t^6-182 t^5+167 t^4-122 t^3+47 t^2-12
   t+1\right)  \, .
\fe
Just as in  the $SU(N)$ case, since the coefficients $b_{SO(2N)}(s)$ are uniquely determined by the perturbation theory results, $B_{SO(2N)}(t)$ is related to the perturbative expression in terms of Laguerre polynomials \eqref{pertsoeven}. This allows us to obtain $B_{SO(2N)}(t)$ for arbitrary $N$ \eqref{eq:SUNnew} by an analysis analogous to that used for $\C_{SU(N)}$, we find, 
\ie \label{eq:SONnew}
B_{SO(2N)}(t)  &= - t \int_0^{\infty} dx \,  e^{-x t} \,   \tilde{B}_{SO(2N)}(x)  \, ,
\fe
with 
\ie \label{eq:SONnew2}
\tilde{B}_{SO(2N)}(x)  &=   { x^{{3\over 2}} \over 2} \partial_x \Big\lbrace  x^{{3\over 2}} \partial_{x} \Big[ e^{-{x}} \sum_{i,j=1}^N \left( L_{2(i-1)}\left({ x}\right)L_{2(j-1)}\left({ x}\right)- L_{2(i-1)}^{2(j-i)}\left({x}\right)L_{2(j-1)}^{2(i-j)}\left({ x}\right) \right) \Big] \Big\rbrace   \, .
\fe
One can easily verify that \eqref{eq:SONnew} reproduces the examples given in \eqref{eq:SON-ex}.

Similarly,  for  $SO(2N+1)$ (or equivalently $USp(2N)$), we find
\bea
B^1_{SO(2N+1)}(t) =B^2_{USp(2N)}(t) =  \frac{\cQ^1_{SO(2N+1)}(t)}{(t+1)^{4N-1}} \, , \qquad  B^2_{SO(2N+1)}(t) =B^1_{USp(2N)}(t) = \frac{\cQ^2_{SO(2N+1)}(t)}{(t+1)^{2N+3}} \, ,
\label{BSOodd2def}
\eea
where $\cQ^1_{SO(2N+1)}(t)$ and $\cQ^2_{SO(2N+1)}(t)$ are degree-$(4N-3)$ and degree-$(2N+1)$ palindromic polynomials, respectively. 
For $N=1$, as previously mentioned, we have $\C_{SO(3)}(\tau,\bar{\tau}) = \C_{SU(2)}(2\tau,2\bar{\tau})$ hence, using \eqref{eq:BSUN}, we deduce
\ie
B^1_{SO(3)}(t) = 0\,,\qquad B^2_{SO(3)}(t) = B_{SU(2)}(t) = \frac{3t(3t^2-10t+3)}{2(t+1)^5}\,,
\fe
consistent with generic expectations \eqref{intcon}.

It turns out that it is relatively simple to determine $\cQ^2_{SO(2N+1)}(t)$. After examining many examples we find a simple expression
\ie \label{eq:B2Q}
\cQ^2_{SO(2N+1)}(t) = \frac{N}{{2}} (2 N+1) t (t-1)^{2 N-2} \left(3 t^2 -(8 N+2) t+3\right) \, ,
\fe
consistent with $\cQ^2_{SO(3)}(t) = \cQ_{SU(2)}(t) $.
Although a general formula for $\cQ^1_{SO(2N+1)}(t)$ is harder to obtain, nevertheless one may compute it in principle for any $N$ either by use of the Laplace-difference equation or  from the perturbative results. The following are two examples of  $\cQ^1_{SO(2N+1)}(t)$,
 \ie \label{eq:B1Q}
\cQ^1_{SO(5)}(t) &= \cQ^2_{SO(5)}(t) = \cQ^1_{USp(4)}(t) = \cQ^2_{USp(4)}(t) = 15 (t-1)^2 t \left(t^2-6 t+1\right) \, , \cr
\cQ^1_{SO(7)}(t) &= 21 t \left(3 t^8-37 t^7+123 t^6-207 t^5+220 t^4-207 t^3+123 t^2-37 t+3\right) \, .
\fe
As in the $SU(N)$ and $SO(2N)$ cases, the functions $B^i_{SO(2N+1)}(t)$ can be obtained from the perturbative expression in terms of sums of Laguerre polynomials  given in \eqref{pertsoodd}.  The term linear in Laguerre polynomials  in  \eqref{pertsoodd} gives the simpler function, $B^2_{SO(2N+1)}(t)$,
\ie \label{eq:B2new}
B^2_{SO(2N+1)}(t)  &= - t  \int_0^{\infty} dx \,  e^{- x t}\,   \tilde{B}^2_{SO(2N+1)}(x)  \, ,
\fe
with\footnote{The reason  $L_{2i-1}\left({ 2x}\right)$ (rather than $L_{2i-1}\left({x}\right)$) arises  in the definition of $\tilde{B}^2_{SO(2N+1)}(x)$ is because $B^2_{SO(2N+1)}(t)$ is associated with $E(s;2\tau, 2\bar \tau)$. }
\ie \label{eq:B2new2}
\tilde{B}^2_{SO(2N+1)}(x)  &=  { x^{{3\over 2}} \over 2} \partial_x \Big[  x^{{3\over 2}} \partial_{x} \Big( e^{ -x } \sum_{i=1}^N L_{2i-1}\left({ 2x}\right)\Big) \Big]   \, ,
\fe
and the term quadratic in Laguerre polynomials leads to $B^1_{SO(2N+1)}(t)$, 
\ie \label{eq:B1new}
B^1_{SO(2N+1)}(t)  &= - t  \int_0^{\infty} dx \,  e^{- x t}\,   \tilde{B}^1_{SO(2N+1)}(x)  \, ,
\fe
with 
\ie \label{eq:B1new2}
\tilde{B}^1_{SO(2N+1)}(t)  &= { x^{{3\over 2}} \over 2}  \partial_x \Big\lbrace  x^{{3\over 2}} \partial_{x} \Big[ e^{-x }\sum_{i,j=1}^N \Big(L_{2i-1}\left(x \right)L_{2j-1}\left( x \right) -L_{2i-1}^{2(j-i)}\left( x \right)L_{2j-1}^{2(i-j)}\left(x \right) \Big)\Big] \Big\rbrace \, . 
\fe
Again, one can verify that \eqref{eq:B2new} and \eqref{eq:B1new}  are in agreement with the expressions given in \eqref{eq:B2Q} and \eqref{eq:B1Q}, respectively. In particular for $N=1$ we can easily see that $B^1_{SO(3)}(t)=0$ from \eqref{eq:B1new} and $B^2_{SO(3)}(t) = B_{SU(2)}(t)$ from \eqref{eq:B2new}.

As described in the introduction, the functions  $B_{G_N}^i  (t)$ obey the inversion and integration conditions, 
\bea
B_{G_N}^i  (t) = t^{-1}\,B_{G_N}^i (t^{-1})\, , \qquad\qquad \int_0^\infty \frac{dt}{\sqrt t}\, B^i_{G_N} (t)=0 \, ,
\eea
as well as the other integral conditions presented in \eqref{intcon}, which  we have checked for many different values of $N$.
As explained in \cite{Collier:2022emf}, both of these conditions are closely related to modularity of the corresponding lattice sum integrals.\footnote{We would like to thank Scott Collier and Eric Perlmutter for clarifications on this issue.}  

Finally, it would be interesting to obtain expressions for $B_{G_N}^i  (t)$ as explicit functions of $N$ for general gauge group $G_N$, analogous to that of $SU(N)$ given in \eqref{polydef}. Such expressions would allow us to perform non-perturbative checks of the relation $\C_{SO(2N)}(\tau,\bar{\tau}) = \C_{USp(-2N)}(-\frac{\tau}{2},-\frac{\bar{\tau}}{2})$, which we have shown is a property of the perturbative expansion.  Although it is difficult to perform the continuation $N \to -N$ using the expressions given in \eqref{eq:SONnew2} and \eqref{eq:B1new2}, we saw earlier that the Laplace-difference equations are perfectly consistent with this relation.

\subsection{Non-perturbative checks}

The coefficients $b_{G_N}^i  (s)$, or equivalently $B_{G_N}^i  (t)$, were designed to reproduce the  perturbative expressions   of the integrated correlators that are determined by localisation.  It is important to verify that the exact expressions given by \eqref{mainres} also give rise to correct non-perturbative instanton contributions. 
We have already shown, using Laplace-difference equations, that $\C_{SO(4)}(\tau,\bar\tau)  = 2\, \C_{SU(2)}(\tau,\bar\tau)$ and   $\C_{SO(6)}(\tau,\bar\tau)  = \C_{SU(4)}(\tau,\bar\tau)$ for any $\tau$, as expected. So here we will consider more general examples. For the one instanton contributions to $SO(n)$ correlators, we find:
\begin{align}
\C^{(1)}_{SO(5)}(\tau,\bar\tau) &=\C^{(1)}_{USp(4)}(\tau,\bar\tau)= e^{2\pi i \tau} \, 20 \Big[ y^2 (8 y+5)-\frac{\sqrt{\pi }}{4}  e^{4 y} y^{3/2} \left(64 y^2+48 y+3\right)
   \text{erfc}\left(2 \sqrt{y}\right) \Big]  \, ,
   \label{usp4}
\end{align}   
   \begin{align}
\C^{(1)}_{SO(7)}(\tau,\bar\tau) &= e^{2\pi i \tau} \,\frac{21}{32} \, \Big[ y^2 \left(512 y^3+2496 y^2+2824
   y+707\right) \\
   & - \, \frac{\sqrt{\pi } }{4} e^{4 y} y^{3/2} \left(4096
   y^4+20480 y^3+24960 y^2+7936 y+317\right)
   \text{erfc}\left(2 \sqrt{y}\right) \Big] \, , \nn
   \end{align}
   \begin{align}
 \C^{(1)}_{SO(8)}&(\tau,\bar\tau) = e^{2\pi i \tau} \, \frac{7}{1024}  \Big[ y^2 \left(45056 y^4+358400 y^3+805632
   y^2+630336 y+136173\right)  \\
   &- \,  \frac{\sqrt{\pi } }{4}{ e^{4 y} y^{3/2} \left(360448
   y^5+2912256 y^4+6792192 y^3+5765760 y^2  +1567800
   y+60435\right) \text{erfc}\left(2
   \sqrt{y}\right)} \Big] \, , \nn    
   \label{so8}
\end{align}
where $y = \pi \tau_2 = 4\pi^2 / g_{_{YM}}^2$.
We have verified that all these results, as well as those with higher $N$, match precisely the one-instanton computation from localisation given in section \ref{sec:yminst}. 

Turning to  $USp(2N)$ we first recall  from \eqref {usp4} that $\C_{USp(4)}(\tau,\bar\tau) = \C_{SO(5)}(\tau,\bar\tau)$.  For higher values of $N$ we have, for example,
\ie \label{eq:USpinst}
\C^{(1)}_{USp(6)}(\tau,\bar\tau)  &= e^{2\pi i \tau} \, 7 \left[ y^2 (8 y+3) (8 y+11)-\frac{ \sqrt{\pi }}{4} e^{4 y} y^{3/2} \left(512
   y^3+960 y^2+360 y+15\right) \text{erfc}\left(2
   \sqrt{y}\right) \right] \, , \cr
 \C^{(1)}_{USp(8)}(\tau,\bar\tau)  &= e^{2\pi i \tau} \,   {3\over 2} \Big[ y^2 \left(512 y^3+1728 y^2+1480 y+279\right) \cr
 & -\frac{\sqrt{\pi }}{4}  e^{4 y} y^{3/2} \left(4096
   y^4+14336 y^3+13440 y^2+3360 y+105\right)
   \text{erfc}\left(2 \sqrt{y}\right) \Big] \, .
\fe
We have verified that the above results, as well as the one-instanton contributions to $\C^{(1)}_{USp(2N)}(\tau,\bar\tau)$ for other values of $N$ deduced from the Laplace-difference equation, again agree with the localisation computation in section \ref{sec:yminst}. 

Furthermore, from examples such as those in \eqref{eq:USpinst} we see that the one-instanton contributions  to $\C_{USp(2N)}(\tau, \bar \tau)$  for general values of $N$  in the weak coupling expansion behave as 
\ie \label{eq:USpN1}
\C^{(1)}_{USp(2N)}(\tau,\bar\tau) \sim  e^{2\pi i \tau} \,  \left( y^{1-N} + O(y^{-N}) \right) \, .
\fe
This property is also evident from the localisation result,  \eqref{kinstsp2N}, which implies
\ie \label{eq:USpN2}
\C^{(1)}_{USp(2N)}(\tau,\bar\tau) &=\frac{1}{4} \Delta_{\tau}\Big[ e^{2\pi i \tau} \left \langle \frac{1}{2}   \prod_{j=1}^N \frac{a_j^2}{a_j^2+2} \right \rangle \Big] \cr
& \sim \Delta_{\tau}[ e^{2\pi i \tau} \,  \left( y^{-N} + O(y^{-N-1}) \right)  ] \sim  e^{2\pi i \tau} \,  \left( y^{1-N} + O(y^{-N}) \right) \, .
 \fe
 The behaviour \eqref{eq:USpN1} implies that  the one-instanton contribution to $\C_{USp(2N)}(\tau,\bar\tau)$ is exponentially suppressed in the large-$N$ expansion. In fact, as we will see in the next section, for $USp(2N)$ all the contributions of odd instanton number are suppressed in the large-$N$ limit.  This is  in agreement with semi-classical instanton calculations based on the ADHM construction in \cite{Hollowood:1999ev}.

 The localisation computations of the $k$-instanton contributions with $k>1$ to $\C_{SO(n)}$ and $\C_{USp(2N)}$  are not as explicitly understood as in the case of $\C_{SU(N)}$, due to complications in obtaining explicit expressions for the $k$-instanton Nekrasov partition functions. In section \ref{sec:yminst} we computed the two-instanton contributions to the  Nekrasov partition function  for the $SO(4)$ and $SO(5)$ theories, and the two- and three-instanton contributions to the  $USp(2N)$ theories for $N\le 5$, using the formulation given in \cite{Billo:2015pjb, Billo:2015jyt}.  We have verified that all these multiple-instanton results, which originate from the localisation for the integrated correlators,  agree with the exact formulae described in this section, and they provide further strong evidence to the validity of our conjecture \eqref{mainres}.

\section{The large-$N$ expansion}
\label{sec:largen}

As in \cite{Dorigoni:2021bvj,Dorigoni:2021guq} we will consider two distinct large-$N$ limits:  one of these is  {a generalisation of} the standard 't Hooft limit {of the $SU(N)$ theory}, in which $g_{_{YM}}^2 N$ is fixed so that $g_{_{YM}}^2 \sim 1/N$, and the contributions of Yang--Mills instantons are  exponentially suppressed.  The other large-$N$ limit is one with finite $g_{_{YM}}^2$, in which instantons contribute and S-duality is manifest.
    
  \subsection{The 't Hooft limit}
  \label{thooftn}
This is the limit in which the correlators have topological expansions reminiscent of 't Hooft's analysis of  $SU(N)$ Yang--Mills theory in the large-$N$ limit \cite{tHooft:1973alw}.   However, the details of our analysis depend rather sensitively on whether $g_{_{YM}}^2 N \ll 1$ or $g_{_{YM}}^2 N \gg 1$.  We will consider each in turn.

\subsubsection*{The weakly coupled 't Hooft limit}
In this large-$N$ limit, the correlator $\C_{G_N} (\tau, \bar \tau)$ is dominated by the perturbative contribution $\C_{G_N}^{pert}(\tau_2)$ \eqref{propseries}, which has an expansion in powers of $a_{G_N}$ (defined in \eqref{pertgroup}), that is given by
\bea
\C_{G_N} (\tau, \bar \tau) \sim  \C_{G_N}^{pert}(\tau_2) \sim c_{G_N} \sum_{g=0}^{\infty} (N_{G_N})^{-g} \,  \C^{(g)}_{G_N} (a_{G_N}) \, , 
\label{topo1}
\eea
 where $c_{G_N}$ is the central charge given in \eqref{confanom} and the parameters $N_{G_N}$  were defined in \eqref{Ngenus}. 

As emphasised in section \ref{pertexpn}, an interesting property of the integrated correlator is that its planar limit is identical for all the gauge groups. Therefore, one may simply use the known all-order planar result of $\C^{(0)}_{SU(N)}$ \cite{Dorigoni:2021guq} and obtain
\ie \label{small-lam}
\C^{(0)}_{G_N} (a_{G_N} ) = \sum_{m=1}^{\infty}  \frac{(-4)^{m+1} \zeta (2 m+1) \Gamma \left(m+\frac{3}{2}\right)^2}{\pi  \Gamma (m)
   \Gamma (m+3)} (a_{G_N})^m  \, .
\fe  
This sum converges for $|a_{G_N}|<\frac{1}{4}$, and one can perform the convergent sum and obtain 
\bea 
\label{small-lam2}
\C^{(0)}_{G_N} (a_{G_N}) = a_{G_N} \int_0^{\infty} dw \,  w^3 \frac{   _1F_2\left(\frac{5}{2};2,4\,\Big\vert-{4 w^2 a_{G_N}
   }\right)}{ \sinh^2(w)} \, ,
\eea
in agreement with \cite{Binder:2019jwn, Dorigoni:2021guq}) and, the results given in \cite{Alday:2021vfb} (after they are simplified).
Sub-leading coefficients for each gauge group,  $C_{G_N}^{(g)}$ with $g\ge 1$, can be determined by using the Laplace-difference equations to any desired order (and agree with the sub-leading terms listed in \cite{Alday:2021vfb} for each gauge group).

\subsubsection*{The strongly coupled 't Hooft limit}
\label{sec:strongthooft}

We now turn to the large-$N$ expansion of   $\C_{G_N}(\tau,\bar\tau)$ in the regime in which 't Hooft coupling is large.   Once again this is  a topological series analogous to \eqref{topo1}.  Whereas in the $SU(N)$ case the 't Hooft coupling is defined by $ \lambda_{SU(N)} :=  g_{_{YM}}^2N$ the holographic connection with superstring amplitudes suggests that the strong-coupling 't Hooft parameter takes a somewhat different form in terms of $N$ in the case of  $\C_{SO(2N)}, \C_{SO(2N+1)}$ and $\C_{USp(2N)}$.
The holographic interpretation for general classical Lie groups  \cite{Blau:1999vz} will be briefly reviewed in appendix~\ref{sec:stringy}  where it will be seen that the  natural definition of the expansion coefficients for the various groups take the form
\bea 
\lambda_{SU(N)}  :=  g_{_{YM}}^2\, N\,, \qquad \lambda_{SO(n)}   :=  {g_{_{YM}}^2 } \left({n \over 2}-{1\over 4} \right) \, , \qquad \lambda_{USp(n)}   :=  {g_{_{YM}}^2 } \left({n \over 2}+{1\over 4} \right)\, ,
\label{eq:lamval}
\eea
which are in accord with  \cite{Alday:2021vfb} and are of the form $\lambda_{G_N}  := g_{_{YM}}^2\, \flux$, where $\flux$ is the $RR$ five-form flux in the appropriate orientifold background given by 
\begin{equation}\label{eq:Fluxes}
 \fluxSU   :=  N\,,\qquad\qquad \,\fluxSO   :=  \frac{n}{2}-\frac{1}{4}\,,\qquad\qquad\,\fluxUSp := \frac{n}{2}+\frac{1}{4}\,,
 \end{equation} 
 as discussed in appendix~\ref{sec:stringy}. 
 We see that, with the exception of the $SU(N)$ case, the  $\lambda_{G_N}$ are different from $a_{G_N}$, which were the expansion parameters relevant in the weak coupling region and defined in \eqref{pertgroup}. 
 
 With these definitions of the parameters we find that in the strong 't Hooft coupling region the large-$\flux$ asymptotic expansion of the integrated correlators takes the following form
\ie  \label{eq:ggg}
\C_{G_N}(\lambda) \sim   \sum_{g=0}^{\infty} (\flux)^{2-2g} f^{(g)}_{G_N}(\lambda_{G_N} ) \, .
\fe
For each {value of $g$  the asymptotic expansion of  $f^{(g)}_{G_N}(\lambda_{G_N} )$ in the large-$\lambda_{G_N}$ limit has the form of
\ie
f^{(g)}_{G_N}(\lambda_{G_N} ) \sim \sum_{\ell} b_{\ell}^{(g)} \lambda_{G_N}^{-\ell/2} \, .
\fe
}The $g=0$ term, $f^{(0)}_{G_N}(\lambda_{G_N})$, can be obtained from \eqref{small-lam2}  by expressing $a_{G_N}$ in terms of   $\lambda_{G_N}$ and expanding for large $\lambda_{G_N}$. 
For $SU(N)$, we simply have  $\lambda_{SU(N)} = 4\pi^2a_{SU(N)}=g_{_{YM}}^2N $.  For the other classical gauge groups the    
the relations between $a_{G_N}$ and $\lambda_{G_N}$ are also simple in the large-$\flux$ limit, where to leading order we have
\ie
\label{lamatrans}
\lambda_{SO(n)} = 2\pi^2 a_{SO(n)} + O(n^{-1}) \, , \qquad \lambda_{USp(n)} = 4\pi^2 a_{USp(n)}  + O(n^{-1}) \,.
\fe  
Using these relations and given that the planar contributions are identical for all gauge groups, the large-$\lambda_{G_N}$ expansions are determined by the $SU(N)$  results \cite{Binder:2019jwn, Dorigoni:2021guq}.  We find  
\bea
f^{(0)}_{{USp(n)}}(\lambda)=  f^{(0)}_{{SO(n)}}(2\lambda) = {1\over 8} +\sum_{m=1}^{\infty}  \frac{2^{1-2m}   \Gamma
   \left(m-\frac{3}{2}\right) \Gamma
   \left(m+\frac{3}{2}\right) \Gamma (2 m+1)\zeta (2 m+1) }{\pi  \,  \Gamma
   (m)^2} \lambda^{-m-{1\over 2}}\, ,
   \label{f0lam1}
\eea
where the factor of 2 in the argument of $ f^{(0)}_{{SO(n)}}(2\lambda)$ originates with \eqref{lamatrans}.
The sub-leading terms (the first few of which were determined in \cite{Alday:2021vfb}) can also be determined in a systematic manner from the Laplace-difference equations.  They have a structure that corresponds to terms that would arise in the low energy expansion of type IIB superstring amplitudes in an $AdS_5\times S^5/\Z_2$ orientifold background.

In  \cite{Dorigoni:2021guq} it was shown that the large-$\lambda$ expansion of  $f^{(g)}_{SU(N)}(\lambda)$ is an asymptotic series, which is not Borel summable. The analysis was carried out for $g=0$ and $g=1$ but in all likelihood it extends to all values of  $g$.   Applying ideas from resurgence similar to \cite{Aniceto:2015rua,Dorigoni:2015dha,Arutyunov:2016etw, Dorigoni:2021guq}, the  large-$\lambda$ expansion of the correlator $\C_{SU(N)}(\tau,\bar\tau)$  therefore receives non-perturbative contributions, which behave as $e^{-\alpha \sqrt{\lambda}}$ for some constant $\alpha$. The same considerations appear in $\C_{SO(n)}$ and $\C_{USp(n)}$. In particular, the  $g=0$ terms in \eqref{f0lam1} take the same form as in $f^{(0)}_{SU(N)}(\lambda)$, and therefore have the same non-perturbative contributions. Similarly, the sub-leading powers of $\flux$ (terms with  with $g>0$ in \eqref{eq:ggg}), have  large-$\lambda$ expansions with  very similar structures for all classical gauge groups.   Once again, they are not Borel summable and are expected to have similar non-perturbative completions. It would be interesting to understand the path-integral semi-classical origin of these non-perturbative corrections, which have a behaviour suggestive of world-sheet instantons  \cite{Dorigoni:2021guq}.

\subsection{The fixed-$g_{_{YM}}^2$ limit}
\label{fixedg}

In  this limit the large-$N$ expansion of the integrated correlator  is manifestly invariant under Montonen--Olive (or GNO) duality
 \cite{Chester:2019jas, Chester:2020vyz, Dorigoni:2021guq, Alday:2021vfb}.  In order to determine an unlimited number of terms in this expansion we will combine the   Laplace-difference equations with the results of the large-$N$ expansion of $\C_{SU(N)}(\tau, \bar \tau)$ determined  in  \cite{Chester:2019jas, Dorigoni:2021guq}, which are summarised up to order $N^{-\frac{11}{2}}$ as follows
 \begin{align} \label{oldexpand}
& \C_{SU(N)}(\tau,\bar\tau) \sim \frac{N^2}{4} - \frac{3N^\half}{2^4}E({\scriptstyle \frac 32}; \tau,\bar\tau)+\frac{45  N^{-\half}}{2^8}E({\scriptstyle \frac 52}; \tau,\bar\tau) \\
&\notag+  {N^{-\frac{3}{2}}}\Big[\frac{4725}{2^{15}} E({\scriptstyle \frac 72}; \tau,\bar\tau)-\frac{39}{2^{13}}E({\scriptstyle \frac 32}; \tau,\bar\tau) \Big] +{N^{-\frac{5}{2}}}\Big[ \frac{99225}{2^{18}} E({\scriptstyle \frac 92}; \tau,\bar\tau)  -\frac{1125}{2^{16}} E({\scriptstyle \frac 52}; \tau,\bar\tau)\Big]\\
&\notag +  {N^{-\frac{7}{2}}}\Big[\frac{245581875}{2^{27}}E({\scriptstyle \frac{11}{2}}; \tau,\bar\tau)  -\frac{2811375}{2^{25}}E({\scriptstyle \frac 72}; \tau,\bar\tau)+\frac{4599}{2^{22}} E({\scriptstyle \frac 32}; \tau,\bar\tau)\Big] \cr
&\notag +  {N^{-\frac{9}{2}}}\Big[\frac{29499294825}{2^{31}}E({\scriptstyle \frac{13}{2}}; \tau,\bar\tau)  -\frac{39590775 }{2^{26}}E({\scriptstyle \frac 92}; \tau,\bar\tau)+\frac{1548855 }{2^{27}} E({\scriptstyle \frac 52}; \tau,\bar\tau)\Big]  \cr
&\notag +  {N^{-\frac{11}{2}}}\Big[\frac{40266537436125}{2^{38}}E({\scriptstyle \frac{15}{2}}; \tau,\bar\tau)  -\frac{397105891875}{2^{36}}E({\scriptstyle \frac {11}{2} }; \tau,\bar\tau)+\frac{2029052025}{2^{34}} E({\scriptstyle \frac 72}; \tau,\bar\tau) \cr
&- \frac{3611751}{2^{32}} E({\scriptstyle \frac 32}; \tau,\bar\tau)\Big]  +O(N^{-\frac{13}{2}})  \, .
\end{align}
As shown in \cite{Dorigoni:2021bvj, Dorigoni:2021guq}, this result can be obtained directly from the large-$N$ expansion of \eqref{eq:SUrevB}. In these references, it was also shown that the Laplace-difference equation, \eqref{lapdiffSUN} imposes strong constraints on the form of  \eqref{oldexpand}.  Thus, once the coefficients of the Eisenstein series with the highest values of $s$  at every power of $1/N$ (the `highest-$s$'  coefficients) are known, the Laplace-difference equation determines all  the remaining expansion coefficients.  But the highest-$s$ coefficients  are completely determined by the planar-limit result obtained from \eqref{small-lam2}, as shown in \cite{Dorigoni:2021guq}. Therefore  the large-$N$ expansion of the correlator is fully determined from \eqref{small-lam2}  and the Laplace-difference equation. 

Let us now consider the large-$N$ expansion of the integrated correlators of the $SO(n)$ theory using the Laplace-difference equation \eqref{lapdiffSO}. We will solve the equation order by order in $1/\fluxSO$, using the input of the large-$N$ expansion of $\C_{SU(N)}$, that was reviewed in the previous paragraph. We begin by making an ansatz for the  the large-$\fluxSO$ expansion of the $\C_{SO(n)}(\tau, \bar\tau)$, 
\ie \label{largelam}
\C_{SO(n)}(\tau, \bar\tau)   \sim (2\fluxSO)^2 \, \tilde{f}_2 (\tau, \bar\tau) + (2\fluxSO) \tilde{f}_1(\tau, \bar\tau) + \tilde{f}_0(\tau, \bar\tau) + \sum_{\ell=0}^{\infty} {(2\fluxSO)}^{{1\over 2} -\ell} \,{f}_{\ell}(\tau, \bar\tau) \, .
\fe
Here we choose to expand $\C_{SO(n)}(\tau, \bar\tau)$ in powers of $2\fluxSO$ in order to make the comparison with the expansion of $\C_{SU(N)}$ in \eqref{oldexpand} clearer.  Substituting the ansatz \eqref{largelam} into the Laplace-difference equation \eqref{small-lam2} and expanding order by order in $1/n$ determines the equations satisfied by the coefficients of the powers of $\fluxSO$.    At  order $n^2$, $n^1$ and $n^0$, the Laplace-difference equation  leads to the conditions 
\ie
\tilde{f}_2 (\tau, \bar\tau) = {1\over 8} \, , \qquad  \Delta_\tau \tilde{f}_1(\tau, \bar\tau)  =\Delta_\tau \tilde{f}_0(\tau, \bar\tau)  =0 \, .
\fe
Invariance under $SL(2,\Z)$ implies  $\tilde{f}_0(\tau, \bar\tau)$ and  $\tilde{f}_1(\tau, \bar\tau)$  must be independent of $\tau$, and are therefore constant.  Comparison with the perturbative expansion shows that these constants must each  vanish. Indeed  $\fluxSO^2  \tilde{f}_2 (\tau, \bar\tau) = \fluxSO^2\, /8$ precisely matches the supergravity expression.

The equations associated with half-integer powers in $n$ are more interesting.  The Laplace-difference equation  \eqref{small-lam2}  implies that each coefficient function $f_\ell(\tau,\bar\tau)$ must satisfy an inhomogeneous Laplace equation.     The first such equation arises at order ${n}^{1\over 2}$ and  takes the following form, 
\ie
 \left( \Delta_\tau +{1\over 4} \right)  {f}_{0}(\tau, \bar\tau) =-  {3 \over 32} E\left( \threeh; \tau, \bar \tau \right) \, .
 \fe
The above equation has the  $SL(2, \Z)$ invariant solution
\ie
{f}_{0}(\tau, \bar\tau) = -{3 \over 32 } E\left( \threeh; \tau, \bar \tau \right)  + \alpha\, E\left( \half; \tau, \bar \tau \right) \, .
\fe
The last term proportional to $E(\half; \tau, \bar\tau)$ is  an arbitrary multiple of the modular invariant solution of the homogeneous equation
\ie
 \left( \Delta_\tau +{1\over 4} \right)  {f}_{0}(\tau, \bar\tau) =0 \, .
\fe
However,  the coefficient $\alpha$ must vanish since the zero mode of $E\left( \half; \tau, \bar \tau \right)$ is proportional to $\tau_2^{\frac{1}{2}} \log(\tau_2)$, which is inconsistent with the known perturbative result.  Likewise, at order $n^{-\half}$  we find the equation is given by
\ie
\left( \Delta_\tau -{3 \over 4} \right)  {f}_{1}(\tau, \bar\tau) = {135 \over512 } E\left( \fiveh; \tau, \bar \tau \right)  \, ,
\fe
which implies 
\ie
{f}_{1}(\tau, \bar\tau)  = {45 \over 512 }E\left( \fiveh; \tau, \bar \tau \right)  + \beta \, E\left( \threeh; \tau, \bar \tau \right)  \, ,
\fe
where $E\left( \threeh; \tau, \bar \tau \right)$ is the modular invariant solution of the homogeneous equation. However,  either by comparing with the perturbative results \cite{Alday:2021vfb} or with the one-instanton contributions presented in appendix \ref{1-instantonN}, we find  $ \beta =0$., so the coefficient of the inhomogeneous equation again vanishes.
At order, $n^{-\threeh}$ we find
\ie
\left( \Delta_\tau -{15 \over 4} \right)  {f}_{2}(\tau, \bar\tau) = {23625 \over 2^{16} } E\left( \sevenh; \tau, \bar \tau \right) +{333 \over 2^{14} } E\left( \threeh; \tau, \bar \tau \right)  \, . 
\fe
The $SL(2,\Z)$-invariant solution to this equation is given by
\ie \label{eq:ff2}
{f}_{2}(\tau, \bar\tau) ={4725 \over 2^{16}} E\left( \sevenh; \tau, \bar \tau \right)  - {111 \over 2^{14}} E\left( \threeh; \tau, \bar \tau \right) + \gamma \, E\left( \fiveh; \tau, \bar \tau \right) \, ,
\fe
where $\gamma \, E\left( \fiveh; \tau, \bar \tau \right)$ is the modular invariant homogeneous solution, which again has to vanish in order to be consistent with the perturbative result or the one-instanton contribution. 

One may proceed in a similar way to obtain the expressions  for ${f}_{\ell}(\tau, \bar\tau)$ for general $\ell$.  For each value of $\ell$, the function $f_\ell(\tau,\bar\tau)$ gets a contribution proportional to $E\left(\ell + {1\over 2}; \tau, \bar \tau \right)$ from the modular invariant solution of a homogeneous Laplace equation. Such a contribution must have vanishing coefficient since it is inconsistent with the structure \eqref{eq:ggg}.   To see this we may substitute the relation $\tau_2 = 2\pi(n-\frac{1}{2}) /\lambda_{SO(n)}$  into the zero mode of $E\left(\ell + {1\over 2}; \tau, \bar \tau \right)$ (the sum of the $\tau_2^{\ell+\half}$ and $\tau_2^{-\ell+\half}$ terms)   to 
 convert to the variables $\fluxSO$ and $\lambda_{SO(n)}$.   It is easy to see that such a contribution behaves as $\fluxSO^{2-g}$ (instead of $\fluxSO^{2-2g}$), which is inconsistent with the general structure given in \eqref{eq:ggg}. In particular, these solutions to the homogeneous equations would lead to perturbative terms proportional to $\fluxSO$, which are not present in the   perturbative computation \cite{Alday:2021vfb}.
 
We therefore conclude that all of the solutions to the homogeneous equations must have vanishing coefficients. 
  This is similar to the systematics of the solution of the Laplace-difference equation of the $SU(N)$ correlator, as analysed in \cite{Dorigoni:2021guq} where, at order  $N^{{1\over 2} -\ell}$ the coefficient multiplying $E\left(\ell + {3\over 2}; \tau, \bar \tau \right)$ was not determined by the Laplace-difference equation. 

Once the solutions to the homogeneous equations have  been set to zero, the Laplace-difference equations determine the coefficients in the large-$\fluxSO$ expansion uniquely.
In this manner we find
\begin{align}
\label{eq:obr}
& 2\, \C_{SO(n)}(\tau,\bar\tau)\sim \frac{(2\fluxSO)^2}{4} - \frac{3(2\fluxSO)^\half}{2^4}E({\scriptstyle \frac 32}; \tau,\bar\tau)+\frac{45  (2\fluxSO)^{-\half}}{2^8}E({\scriptstyle \frac 52}; \tau,\bar\tau) \\
&\notag+ { (2\fluxSO)^{-\frac{3}{2}}}\Big[\frac{4725}{2^{15}} E({\scriptstyle \frac 72}; \tau,\bar\tau)-\frac{111}{2^{13}}E({\scriptstyle \frac 32}; \tau,\bar\tau) \Big] + { (2\fluxSO)^{-\frac{5}{2}}}\Big[ \frac{99225}{2^{18}} E({\scriptstyle \frac 92}; \tau,\bar\tau)  -\frac{3825}{2^{16}} E({\scriptstyle \frac 52}; \tau,\bar\tau)\Big]\\
&\notag +  { (2\fluxSO)^{-\frac{7}{2}}}\Big[\frac{245581875}{2^{27}}E({\scriptstyle \frac{11}{2}}; \tau,\bar\tau)  -\frac{10749375}{2^{25}}E({\scriptstyle \frac 72}; \tau,\bar\tau)+\frac{40239}{2^{22}} E({\scriptstyle \frac 32}; \tau,\bar\tau)\Big] \cr
&\notag + {(2\fluxSO)^{-\frac{9}{2}}}\Big[\frac{29499294825}{2^{31}}E({\scriptstyle \frac{13}{2}}; \tau,\bar\tau)  -\frac{164614275  }{2^{26}}E({\scriptstyle \frac 92}; \tau,\bar\tau)+\frac{18332055  }{2^{27}} E({\scriptstyle \frac 52}; \tau,\bar\tau)\Big]  \cr
&\notag +  {(2\fluxSO)^{-\frac{11}{2}}}\Big[\frac{40266537436125}{2^{38}}E({\scriptstyle \frac{15}{2}}; \tau,\bar\tau)  -\frac{1758611806875 }{2^{36}}E({\scriptstyle \frac {11}{2} }; \tau,\bar\tau)+\frac{28855523025}{2^{34}} E({\scriptstyle \frac 72}; \tau,\bar\tau)\cr
&- \frac{103062039}{2^{32}} E({\scriptstyle \frac 32}; \tau,\bar\tau)\Big] +O(\fluxSO^{-\frac{13}{2}}) \, .
   \end{align}
   This expression applies to $\C_{SO(n)}$ for both $n=2N$ and $n=2N+1$.  As described earlier, we have presented the expansion as a series in $(2 \fluxSO)^{-1}$ in order to emphasise similarities in the coefficients with those of the expansion in the $SU(N)$ case, \eqref{oldexpand}.  Indeed, the highest-$s$  terms in the large-flux number expansion are identical for $\C_{SO(n)}(\tau,\bar\tau)$ and $\C_{SU(N)}(\tau,\bar\tau)$, apart from an overall factor of two. As we saw earlier, the coefficients of the Eisenstein series with highest index $s$ are determined by the planar limit. We also know that the planar contributions to the integrated correlators are identical to all gauge groups. These statements imply that the highest-$s$  terms are the same for all gauge groups.\footnote{The overall factor of $2$ is due to the fact that  $c_{SU(N)} \sim \frac{1}{4}\, \fluxSU$, while $c_{SO(n)} \sim \frac{1}{8} \, \fluxSO$ in the large-$\flux$ limit.}  If one re-expands \eqref{eq:obr}  in powers of $c_{SO(n)}^{-1}$ instead of $\fluxSO^{-1}$ the  expansion agrees with the  expressions in \cite{Alday:2021vfb}, which were computed up to $O(c_{SO(n)}^{-{7/4}} )$.  However, using the Laplace-difference equation makes it  easy to obtain the expansion to any desired order.

We have also solved the Laplace-difference equation \eqref{lapdiffUSp} for the coefficients in the large-$\fluxUSp$ expansion of   $\C_{USp(n)}(\tau,\bar\tau)$. At each order in $1/\fluxUSp$ the equation for $\C_{USp(n)}$ is identical to that of  $\C_{SO(n)}$, except that the terms involving the $SU(N)$ correlators depend on the rescaled coupling, $(\tau, \bar \tau) \to (2\tau, 2\bar \tau)$.   Therefore, we find the result is identical to that of the  $SO(n)$ theory, but with $(\tau, \bar \tau) \to (2\tau, 2\bar \tau)$, and  with  $\fluxSO \to \fluxUSp$, so that, in the large-$\flux$ expansion,  
\ie
 \C_{USp(n)} (\tau, \bar \tau) \sim \C_{SO(n)} (2\tau, 2\bar \tau) \Big{|}_{\fluxSO \rightarrow \fluxUSp}\, .
 \label{eq:Spnstrong}
\fe
The rescaling $(\tau, \bar \tau) \to (2\tau, 2\bar \tau)$ in this expression also implies that odd instanton number terms do not contribute to the integrated correlator of $USp(2N)$ in the large-$N$ expansion. In particular, the one-instanton contribution is suppressed, as we showed in \eqref{eq:USpN2} from the explicit one-instanton computation based on localisation.  One can also see the suppression of the odd-number instantons from the general expression of the integrated correlator given in \eqref{eq:SOodd}, 
\begin{align}  \label{eq:USp22}
\C_{USp(2N)}(\tau,\bar\tau) =    \sum_{(m,n) \in \Z^2}  \int_0^\infty dt\left(B^1_{USp(2N)}(t) e^{-t \pi\frac{ |m+n\tau|^2}{\tau_2}}+B_{USp(2N)}^2 (t) e^{-t\pi \frac{ |m+ 2 n \tau|^2}{2 \tau_2}}\right) \, ,
\end{align} 
and recall using \eqref{BSOodd2def} that
\bea
 B^1_{USp(2N)}(t) =  B^2_{SO(2N+1)}(t) =  N (2 N+1) {t \left(3 t^2 -(8 N+2) t+3\right)  \over 2(t-1)^2 (t+1)^3}  \left({t-1 \over t+1} \right)^{2N}\, .
 \label{BUSp1def}
 \eea
Following a similar analysis to that given in \cite{Dorigoni:2021guq}, one can see that in the large-$N$ limit, the contribution to \eqref{BUSp1def} from $B^1_{USp(2N)}(t)$ is  a coupling-independent constant, with corrections that are exponentially suppressed.  This can be seen as follows.  The $k$-instanton contribution arising from $B^1_{USp(2N)}(t)$ can be expressed via a Poisson summation in the form
\ie \label{eq:CC1}
 e^{2\pi i k {\tau_1}} \sqrt{\tau_2}  \sum_{\hat{m}\neq 0, n\neq 0}  \int_0^\infty {dt\over \sqrt{t} } B^1_{USp(2N)}(t) e^{- \hat{m}^2 \pi \tau_2/t - n^2 \pi \tau_2 t} \, , 
\fe
with $\hat{m} n =k$.   This is suppressed because the last factor in  \eqref{BUSp1def}  satisfies $\left({t-1 \over t+1} \right)^{2N} <1$ in the integration region $0< t < \infty$, apart from the boundaries at $t=0$ and $t = \infty$ (which are however are also suppressed due to the exponential terms $e^{- \hat{m}^2 \pi \tau_2/t}$  and $e^{ - n^2 \pi \tau_2 t}$  in \eqref{eq:CC1}, respectively). Similarly, one can show that the perturbative (i.e. zero-instanton)  contribution is also  exponentially suppressed in the large-$N$ limit apart from a coupling independent   constant.

Therefore, only the second term in \eqref{eq:USp22}  survives in the large-$N$ expansion  (apart from the coupling-independent constant mentioned above).  This means that $\C_{USp(2N)}(\tau,\bar\tau)$ only gets contributions from terms with an even number of instantons, which is in accord with the calculation in \cite{Hollowood:1999ev} of the leading $k$-instanton contribution to the large-$N$ limit based on the ADHM construction. Here we see this is true to all orders in large-$N$ expansion.   Using \eqref{eq:SOodd} and \eqref{eq:USpn}, and the analysis discussed above, we find that
\ie
\C_{USp(2N)}(\tau, \bar \tau)  \sim \C_{SO(2N+1)}(2\tau, 2\bar \tau) \, .
\fe
This is in agreement with our earlier findings \eqref{eq:Spnstrong} since $\tilde{N}_{SO(2N+1)} = \tilde{N}_{USp(2N)}$.

The structure of \eqref{oldexpand}, \eqref{eq:obr} and \eqref{eq:Spnstrong} extend  the $SU(N)$  results in \cite{Chester:2019jas}  and the $SO(n)$ and $USp(n)$ results in  \cite{Alday:2021vfb}.  A notable feature of the structure of these large-$N$ expressions is the fact that the Eisenstein series that arise at each order in $1/\flux$ have half-integer index, whereas those that arise at finite $N$ in \eqref{eisenform}  have integer index.   The low order terms in the large-$N$ expressions have a close connection to corresponding BPS terms in the low energy expansion of the holographically dual type IIB superstring amplitudes, as described in the earlier references.

\section{Discussion }
\label{sec:discussion}

In this paper we have proposed  a lattice sum  representation of the integrated correlator, $\C_{G_N}(\tau,\bar\tau)$,  of four superconformal primary operators in the stress tensor multiplet in $\cN=4$ SYM that are defined by \eqref{firstmeasure}  with any classical gauge group.   This generalises the expression proposed for $SU(N)$ gauge groups in  \cite{Dorigoni:2021bvj,Dorigoni:2021guq}.
 Such  integrated correlators, which are determined by supersymmetric localisation,  are highly constrained by maximal supersymmetry and satisfy  a fascinating interplay of properties that reflects the constraints imposed by GNO duality.

 There are several obvious directions in which these ideas could be extended. 
  A challenging objective would be to extend the discussion in this paper to $\cN=4$ SYM with exceptional gauge groups.  These are theories that are self-dual under the action of GNO S-duality.  With gauge groups  $E_6$, $E_7$ and $E_8$, which are simply-laced, the duality group is  $SL(2,\Z)$.   However, the  duality groups in the non simply-laced cases, $G_2$ and $F_4$, are Hecke groups, which have novel features that will not be reproduced in terms of non-holomorphic Eisenstein series.  Since supersymmetric localisation is ill-understood for exceptional groups an alternative procedure is needed, perhaps making use of the modular  anomaly equation, as suggested in \cite{Billo:2016zbf}.  Another challenge is to construct expressions for integrated $n$-point correlators with $n>4$.  Although the general problem is daunting,  following the methods of this paper and the previous results of \cite{Green:2020eyj, Dorigoni:2021rdo}, it should be possible to obtain exact expressions for integrated maximal $U(1)_Y$-violating   $n$-point correlators with $n>4$ for all classical gauge groups, which  transform covariantly under GNO S-duality (where $U(1)_Y$ is the bonus symmetry \cite{Intriligator:1998ig}). 
 
   Another interesting direction is to formulate lattice representations for other integrated correlators.  In particular,  the correlator given by $\partial_m^4 \log Z(m,\tau,\bar\tau)|_{m=0}$ was analysed in the large-$N$ limit for $SU(N)$ gauge groups, in \cite{Chester:2020vyz,Chester:2020dja}.  In that case the coefficients of integer powers of $1/N$ are generalised Eisenstein series that satisfy inhomogeneous Laplace eigenvalue equations with sources terms that are quadratic in non-holomorphic Eisenstein series. It would be of interest to discover the structure of such correlators at finite values of $N$, perhaps using the recent results of \cite{Dorigoni:2021jfr,Dorigoni:2021ngn}, and for more general gauge groups.
   
  More generally, it should be of interest to consider integrated correlators in a wider context.
  While integration over the operator insertion points obviously averages over the detailed form of any correlator, it remains uncertain as to how much information may be retrieved by considering the set of all possible integrated correlators. 
 Clearly, supersymmetry has played a crucial r\^ole in constraining the integrated correlators we have considered, so it would be interesting to understand how deformations that break supersymmetry affect their structure.  Finally, it would be of interest to understand the extent to which properties of the integrated correlators can be used as probes of the fundamental structure of string theory.

%%%%% 
 \section*{Acknowledgements}
 %%%%%
 
 We would like to thank Shai Chester, Stefano Cremonesi, Scott Collier, Nick Dorey, Francesco Fucito, I\~naki Garc\'ia-Etxebarria, Amihay Hanany, Francisco Morales and Eric Perlmutter for useful conversations.   CW is supported by a Royal Society University Research Fellowship No. UF160350.  
%%%%%%%%%%%%%%%%%%%%%

\appendix
\section{Goddard--Nuyts--Olive duality}
\label{secGNO}
The 1977 paper by  Goddard, Nuyts and Olive \cite{Goddard:1976qe} showed that while electric charges in gauge theories take their values in the weight lattice of the gauge group $G$, the magnetic charges take their values in the  lattice of a dual group $^L{G}$.  Table~\ref{table1} lists the dual groups corresponding to each of the classical Lie groups. Montonen and Olive  \cite{Montonen:1977sn}, conjectured  that there is a duality that identifies a gauge theory with gauge group $G$  and coupling $g_{_{YM}}$ with a theory with gauge group $^LG$ and coupling $^Lg_{_{YM}}=4\pi/g_{_{YM}}$.  The r\^oles of electric and magnetic charges are interchanged by this duality.   It was later understood \cite{Witten:1978mh}  that such a duality requires supersymmetry, and in 1979 it was argued \cite{Osborn:1979tq} that this duality could be realised in $\cN=4$ SYM in which the $\Z_2$ inversion of the coupling is naturally extended to $SL(2,\Z)$  acting on the complex coupling $\tau=\frac{\theta}{2\pi}+i\frac{4\pi}{g_{_{YM}}^2}$  and the spectrum contains infinite towers of dyonic states carrying both electric and magnetic charge.
\renewcommand{\arraystretch}{1.5} 
\begin{table}[htp]
\begin{center}
\begin{tabular}{|c|c|}
$G_N$&$^L{G_N}$\cr
\hline
$U(N)$  &  $U(N)$\cr
$SU(N)$&$PSU(N) = SU(N)/\Z_N$\cr
$Spin(2N)$ &$SO(2N)/\Z_2$\cr
$Sp(N) = USp(2N)$  &   $SO(2N+1)$\cr
$Spin(2N+1)$ & $Sp(N)/\Z_2= USp(2N)/\Z_2$\cr
$G_2$ & $G_2$ \cr
$F_4$ & $ F_4$\cr
$E_{r=6,7,8}$ & $ E_{r}/\Z_{9-r}$\cr
\hline
\end{tabular}
\end{center}
\caption{\label{table1} Langlands/GNO relation between classiical Lie groups and their dual groups.}
\end{table}%

This story has close connections to the Langlands programme \cite{Kapustin:2006pk} and the GNO dual group is identified with the Langlands dual group (hence the superscript on $^LG$).  The extensive connections between the geometric Langlands programme and the dualities of $\cN=4$ SYM is explored in \cite{Kapustin:2006pk} and subsequent papers.

The integrated correlators that are the subject of this paper are not sensitive to the discrete stability groups of $^LG$ listed in the right-hand column of table~\ref{table1}.  They also do not distinguish between  $Spin(N)$ and $SO(N)$.   This means that the  discussions in this paper are at the level of the Lie algebra, $\mathfrak{g}_N$, as shown by the labels  in table~\ref{table2} .  

The S-duality transformation maps a theory with gauge group $G$ into one with gauge group $^LG$. 
 The Montonen--Olive  inversion of the coupling constant, $\tau_2 \to \tau_2^{-1}$  generalises to the  $\hat S$ and $T$ transformations, which are defined by 
\bea
\label{sduality}
&& T\, :\, (G_N,\tau) \to (G_N,  \tau+1)\,, \nn\\
&&\hat S\, :\, (G_N,\tau) \to (^LG_N, -\frac{1}{r\tau})\,,
\eea
where $r$ is the square of the ratio of the long and short roots of the Lie algebra of $G_N$.   In the simply laced cases, i.e. $SU(N)$ and  $SO(2N) $, $r=1$ and $\hat S \equiv S: \tau\to - 1/\tau$ reduces to  the Montonen--Olive transformation when $\tau_1=0$.  In these cases $S$ and $T$ are the generators of the  discrete self-duality group $SL(2,\Z)$, under which
\bea
 \tau  \underset {SL(2,\Z)} \to  \frac{a\tau+b}{c\tau+d} \,,
 \label{sl2t}
\eea
where $a,b,c,d\in \Z$ with $ad-bc=1$.

\renewcommand{\arraystretch}{1.8}
\begin{table}[htp]
\begin{center}
\begin{tabular}{|c|c|}
$\mathfrak{g}_N$&$^L{\mathfrak{g}_N}$\cr
\hline
$su(N)$&$su(N)$\cr
$so(2N)$ &$so(2N)$\cr
$usp(2N)$  &   $so(2N+1)$\cr
$so(2N+1)$ & $usp(2N)$\cr
\hline
\end{tabular}
\end{center}
\caption{\label{table2} Duality relations of relevance to this paper.}
\end{table}%

 In the non simply-laced cases of interest to us  $r=2$ and  the $\hat S$ transformation $\tau\to -1/(2\tau)$ maps theories with gauge groups  $SO(2N+1)$ and $USp(2N)$ into each other.\footnote{This ratio is $r=1$ for the exceptional groups  $E_6,E_7$ and $E_8$, while $r=2$ for $F_4$, and $r=3$ for $G_2$. All the exceptional groups are self-dual (possibly modulo some discrete quotient), i.e. $^LG = G$. when $G= G_2, F_4, E_6, E_7,E_8$.}   In these cases $\hat S$ generates an $SL(2,\RR)$ transformation that is not in $SL(2,\Z)$. It is easy to see that the operators $\hat S T\hat S$ and $T$ generate a $\Gamma_0(r)$ subgroup of $SL(2,\Z)$ (which is a subgroup in which $c=0\,{\rm mod}\, r$).  In other words $\Gamma_0(2)$ is a self-duality group that maps $\C_{G_N}$ into $\C_{G_N}$ and $\C_{^LG_N}$ into $\C_{^LG_N}$.   
  
 There are a number of distinctive features involved in S-duality for gauge theories with exceptional groups \cite{Kapustin:2006pk, Argyres:2006qr}.   In the simply-laced cases ($E_6$, $E_7$ and $E_8$). S-duality is a symmetry associated with the action of $SL(2,\Z)$.  In the non simply-laced cases ($F_4$ and $G_2$, which have $r=2$ and $r=3$, respectively) S-duality is again a symmetry, but the presence of both long and short roots implies that the duality group is a Hecke group rather than a subgroup of $SL(2,\Z)$, which is  generated by $\hat S$ and $T$.

\section{Integrated correlators from localisation}
\label{pertlocal}

In this appendix, we will review the computation of the integrated correlators \eqref{firstmeasure} using supersymmetric localisation.  We begin with a brief review of the application of localisation to the calculation of integrated correlators.
 
\subsection{Review of  integrated correlators}
\label{pestunrev}

The starting point is the partition function of $\cN=2^*$ SYM  on $S^4$,  which  
was determined by Pestun  using supersymmetric localisation in  \cite{Pestun:2007rz}, where it was shown to have the form\footnote{The subscript on $Z_{G_N}$ indicates that the gauge group is $G_N$ .}
\bea
 Z_{G_N}(m,\tau,\bar\tau)  &=&  \frac{1}  {\cN_{G_N}} \int d^r a\, v_{G_N}(a) \, e^{-\frac{8\pi^2}{g_{_{YM}}^2} \langle a,a\rangle } \, \hat Z_{G_N}^{pert}(m,a) \,  |\hat Z_{G_N}^{inst}  (m ,\tau,a)|^2\nn\\
   &=& \langle \,   Z_{G_N}^{pert}(m,a) \,  |\hat Z_{G_N}^{inst}  (m ,\tau,a)|^2  \,  \rangle_{G_N}\,,
 \label{partfun}
 \eea
 where the integration variable $a$ runs over the $r$-dimensional Cartan subalgebra of $G_N$, $v_{G_N}(a)$ is the Vandermonde determinant associated with the group $G_N$,  and  the Killing form $\langle a,a\rangle$ is equal to $\tr_s (a\,a) \,/(2 T_s)  $, where $T_s$ is  the Dynkin index and $s$ denotes the representation. The normalisation factor $\cN_{G_N}$ is given by
 \ie
\cN_{G_N}=  \int d^r a\, v_{G_N}(a) \, e^{-\frac{8\pi^2}{g_{_{YM}}^2} \langle a,a\rangle } \, .
 \fe

    We see from   \eqref{partfun}   that the expectation value of a general function $F(a_i)$ is defined by
 \bea
 \langle F(a_i)\rangle_{G_N} =  \frac{1}{\cN_{G_N}} \int d^r a\, v_{G_N}(a) \, e^{-\frac{8\pi^2}{g_{_{YM}}^2} \langle a,a\rangle } \, F(a_i)\,,
 \label{measure}
 \eea
 so that with the given definition for $\cN_{G_N}$ above, we have $\langle \,   1  \,  \rangle_{G_N}=1$. 
 
 The perturbative contribution to the partition function is one-loop exact and is given by the classical factor proportional to $\exp(-8 \pi^2 \langle a,a\rangle/g_{_{YM}}^2)$ multiplying the one-loop term,
 \bea
 \hat Z_{G_N}^{pert}(m,a) =  \frac{1}{H(m)^r} \prod_{\alpha\in \Delta} \frac{ H(\alpha\cdot a) }  {\big[ H(\alpha \cdot a+ m)H(\alpha \cdot a- m)  \big]^{\frac{1}{2}}  }\,.
\label{zpert}
\eea
Here $r$ denotes the rank of $G_N$, while the product runs over the set of roots.
The function $H(z)$ is given by $H(z)=e^{-(1+\gamma)z^2}\, G(1+iz)\, G(1-iz)$, where  $G(z)$ is Barnes G-function (and $\gamma$ is the Euler constant).  
    The factor of  $|\hat Z_{G_N}^{inst}|^2 = \hat Z_{G_N}^{inst} \,\hat{\bar Z}_{G_N}^{inst}$  in \eqref{partfun} is the contribution from the Nekrasov partition function and describes the contributions from instantons and anti-instantons localised at the north and south  poles of $S^4$.

     The integrated correlation functions of interest for the present paper were defined  in \cite{Binder:2019jwn}  (for $G_N = SU(N)$)  where they were obtained by acting on $\log Z_{G_N}$ with various derivatives with respect to  the hypermultiplet mass, $m$, and the complex coupling, $\tau$, followed by the limit $m\to 0$, as displayed in \eqref{firstmeasure}.  In the same reference \cite{Binder:2019jwn}, it was shown that this quantity is equal to the correlator of four superconformal primary operators  of the stress tensor supermultiplet integrated over their positions with a specific measure that maintains supersymmetry.

The result may be separated into perturbative and instanton contributions since
\ie
\partial^2_{m} \log Z_{G_N}  \big{|}_{m=0} = \partial^2_{m} \log Z_{G_N}^{pert}  \big{|}_{m=0} + \partial^2_{m} \log Z_{G_N}^{inst}  \big{|}_{m=0} \, ,
\fe
where each contribution can be expressed as an expectation value in a gaussian matrix model,   
\ie
 \partial^2_{m} \log Z_{G_N}^{pert}  \big{|}_{m=0}  = \langle \partial_m^2 \hat Z_{G_N}^{pert} \big{|}_{m=0} \rangle_{G_N}\, , \quad 
  \partial^2_{m} \log Z_{G_N}^{inst}  \big{|}_{m=0}  = \langle \partial_m^2 \hat Z_{G_N}^{inst} \big{|}_{m=0} \rangle_{G_N} \, .
 \fe
 The gaussian model expectation value, $\langle \dots \rangle_{G_N}$, is defined by \eqref{measure} and its explicit form for each gauge group is 
 given in appendix~\ref{pertcont}, where the expressions  for the perturbative parts of $\C_{G_N}$ determined in \cite{Alday:2021vfb} are reviewed.  A review of the general structure of the instanton contributions that were discussed in   \cite{Billo:2015pjb, Billo:2015jyt}, is given in appendix~\ref{instcont}.

 \subsection{Perturbative contributions}
 \label{pertcont}
 The discussion in  \cite{Alday:2021vfb} focussed on the perturbative sector, where  $\hat Z_{G_N}^{inst}=1$, and where the partition function has no dependence on $\tau_1=\theta/(2\pi)$. 
In this subsection we will review the explicit form of this measure,  as well as the expressions for $\hat Z_{G_N}^{pert}$, given in  \cite{Alday:2021vfb}  for each classical gauge group. 
  
 \begin{itemize}
 \item $SU(N)$
 \ie
 \hat Z_{SU(N)}^{pert} (m, a_i)= {1\over H(m)^{N-1}} \prod_{i <j } {H^2(a_{ij}) \over H(a_{ij}+m)H(a_{ij} - m)} \, ,
 \fe
 where $a_{ij} = a_i-a_j$.
The expectation value of any function $F(a_i)$ in the $SU(N)$ case is obtained from \eqref{measure} and has the form
\ie \label{SUexp}
\langle  F(a_i) \rangle_{SU(N)}  ={1\over \mathcal{N}_{SU(N)}} \int d^N a \, \delta \big(\sum_i a_i \big)  \prod_{i<j} a_{ij}^2 \, e^{- {8\pi^2 \over g^2_{_{YM}}} \sum_i a_i^2} \, F( a_i ) \, .
\fe

 \item $SO(2N)$
 \ie
\hat Z_{SO(2N)}^{pert} (m, a_i) = {1\over H(m)^{N}} \prod_{i <j } {H^2(a_{ij}) H^2(a^+_{ij}) \over H(a_{ij}+m)H(a_{ij} - m)H(a^+_{ij}+m)H(a^+_{ij} - m)} \, ,
 \fe
where $a_{ij}^+=a_i+a_j$.  The expectation value of $F(a_i)$ in the $SO(2N)$ case is given by the integral
\ie \label{SO1exp}
\langle F(a_i)  \rangle_{SO(2N)}  ={1\over \mathcal{N}_{SO(2N)}} \int d^N a \,  \prod_{i<j} a_{ij}^2 (a^+_{ij})^2 \, e^{- {8\pi^2 \over g^2_{_{YM}}} \sum_i a_i^2} \, F( a_i ) \, .
\fe

 \item $SO(2N+1)$
 \ie
\hat  Z_{SO(2N+1)}^{pert} (m, a_i) &= {1\over H(m)^{N}} \prod_{i}  {H^2(a_{i}) \over H(a_{i}+m)H(a_{i} - m) } \cr
 &  \prod_{i <j } {H^2(a_{ij}) H^2(a^+_{ij}) \over H(a_{ij}+m)H(a_{ij} - m)H(a^+_{ij}+m)H(a^+_{ij} - m)} \, .
 \fe
The expectation value of $F(a_i)$ in the $SO(2N+1)$ case is given by the integral
\ie \label{SO2exp}
\langle   F(a_i) \rangle_{SO(2N+1)}  ={1\over \mathcal{N}_{SO(2N+1)}} \int d^N a \, \prod_{i} a_i^2\,   \prod_{i<j} a_{ij}^2 (a^+_{ij})^2 \, e^{- {8\pi^2 \over  (\delta_{N,1}+1)g^2_{_{YM}}} \sum_i a_i^2} \, F( a_i ) \, .
\fe

 \item $USp(2N)$
 \ie
 \hat Z_{USp(2N)}^{pert} (m, a_i) &= {1\over H(m)^{N}} \prod_{i}  {H^2(2a_{i}) \over H(2a_{i}+m)H(2a_{i} - m) } \cr
 &  \prod_{i <j } {H^2(a_{ij}) H^2(a^+_{ij}) \over H(a_{ij}+m)H(a_{ij} - m)H(a^+_{ij}+m)H(a^+_{ij} - m)} \, .
 \fe
The expectation value of $F(a_i)$ in the $USp(2N)$ case is given by the integral
\ie \label{USpexp}
\langle  F(a_i) \rangle_{USp(2N)}  ={1\over \mathcal{N}_{USp(2N) }} \int d^N a \, \prod_{i} a_i^2\,   \prod_{i<j} a_{ij}^2 (a^+_{ij})^2 \, e^{- {16\pi^2 \over  g^2_{_{YM}}} \sum_i a_i^2} \, F( a_i ) \, .
\fe
 
 \end{itemize}
 
 The perturbative contributions to  $\C_{G_N}(\tau,\bar\tau)$ form an essential ingredient in our discussion.  They are given by substituting the above expressions into \eqref{firstmeasure},  which leads to the following expressions that are given in equation (3.8) of \cite{Alday:2021vfb},\footnote{
The $SU(N)$ case was determined in  \cite{Chester:2019pvm}. Furthermore, the expressions in  \cite{Alday:2021vfb} have been multiplied by a factor of 4 to accord with our conventions.}
\ie
\C_{SU(N)}^{pert} (y) &=-\int_0^\infty d  \omega  \frac{\omega } {2\sinh^2\omega} y^2  \partial_y^2\sum_{i,j=1}^N e^{-\frac{\omega^2}{y }}\Bigg[L_{i-1}\left({ \frac{\omega^2}{y}}\right)L_{j-1}\left({ \frac{\omega^2}{y}}\right)  \\
&\qquad\qquad\qquad\qquad\qquad\qquad\qquad-(-1)^{i-j}L_{i-1}^{j-i}\left({ \frac{\omega^2}{y}}\right)L_{j-1}^{i-j}\left({ \frac{\omega^2}{y}}\right)\Bigg] \, ,    \label{pertSUn} 
\fe
\ie
\C_{SO(2N)}^{pert} (y) &= -\int_0^\infty  d \omega   \frac{\omega }{\sinh^2\omega} y^2  \partial_y^2\sum_{i,j=1}^N e^{-\frac{\omega^2}{y }}\Bigg[L_{2(i-1)}\left({ \frac{\omega^2}{y}}\right)L_{2(j-1)}\left({ \frac{\omega^2}{y}}\right)  \\
& \qquad\qquad\qquad\qquad\qquad\qquad\qquad-L_{2(i-1)}^{2(j-i)}\left({ \frac{\omega^2}{y}}\right)L_{2(j-1)}^{2(i-j)}\left({ \frac{\omega^2}{y}}\right)\Bigg]\,, \label{pertsoeven} 
\fe
\ie
\C_{SO(2N+1)}^{pert} (y)  &= -\int_0^\infty  d \omega   \frac{\omega }{\sinh^2\omega} y^2 \partial_y^2\Bigg \{ e^{-\frac{\omega^2}{y }}\sum_{i,j=1}^N\Bigg[L_{2i-1}\left({ \frac{\omega^2}{y}}\right)L_{2j-1}\left({ \frac{\omega^2}{y}}\right)  \\
& \qquad-L_{2i-1}^{2(j-i)}\left({ \frac{\omega^2}{y}}\right)L_{2j-1}^{2(i-j)}\left({ \frac{\omega^2}{y}}\right)\Bigg]+e^{-\frac{\omega^2 }{2y }}\sum_{i=1}^N L_{2i-1}\left({ \frac{\omega^2}{y}}\right)\Bigg \} \,, \label{pertsoodd} 
\fe
\ie
\C_{USp(2N)}^{pert} (y) &= -\int_0^\infty  d \omega  \frac{\omega }{\sinh^2\omega} y^2 \partial_y^2\Bigg \{ e^{-\frac{\omega^2 }{2y }}\sum_{i,j=1}^N\Bigg[L_{2i-1}\left({\frac{\omega^2 }{2y}}\right)L_{2j-1}\left({\frac{\omega^2 }{2y}}\right)  \\
&\qquad-L_{2i-1}^{2(j-i)}\left({\frac{\omega^2 }{2y}}\right)L_{2j-1}^{2(i-j)}\left({\frac{\omega^2 }{2y}}\right)\Bigg]+e^{-\frac{\omega^2}{y }}\sum_{i=1}^N L_{2i-1}\left({\frac{2\omega^2}{y}}\right)\Bigg \} \,,  \label{pertUSpN} 
\fe
where $y =\pi \tau_2=4\pi^2/g^2_{_{YM}}$, and $L_n^\alpha (x)$ are generalized Laguerre polynomials.\footnote{Laguerre polynomials have previously appeared in the perturbative sector of Wilson  loop calculations in these theories \cite{Fiol:2014fla}.} When $\alpha=0$ one recovers the standard Laguerre polynomials, $L_n(x) := L_n^0(x)$.  For any fixed value of $N$ the above expressions can be expanded in powers of $g_{_{YM}}^2$  to generate the perturbation expansions shown in \eqref{pertexp}.

We note the following:
\begin{itemize}
\item
The $SO(2N+1)$ result only holds for $N>1$ and  the $SO(3)$ case is special since the Dynkin index of $SO(n)$ is discontinuous as $n=3$ is changed to $n>3$.
For $SO(3)$ we must rescale the coupling constant inside the square brackets in  \eqref{pertsoodd}  by $g_{_{YM}}\to\sqrt{2}g_{_{YM}}$. With this rescaling the correlator $\C_{SO(3)}$ is identical to $\C_{SU(2)}$.
\item
These formulae satisfy the isomorphisms $SU(2)\cong SO(3)\cong USp(2)$, $SU(4)\cong SO(6)$, $SO(4)\cong SU(2)\times SU(2)$, and $SO(5)\cong USp(4)$.

\end{itemize}

\subsection{Instanton contributions}
\label{instcont}

 Much of this section is a review of \cite{Billo:2015pjb, Billo:2015jyt}.
 The Nekrasov partition functions that describe the instanton contributions are expressed as infinite Fourier sums,
 \ie
 \hat Z_{G_N}^{inst} (m, \tau, a_{i}) = \sum_{k=0}^{\infty} e^{2\pi i k \tau}  \hat Z^{(k)}_{G_N}(m, a_{i}) \, ,
 \label{instsum}
 \fe   
 where $k$ is the number of instantons, and $\hat Z^{(k)\, inst}_{G_N}(m, a_{i})$ can be conveniently expressed as a contour integral, 
 \ie \label{eq:othergroup}
 \hat Z^{(k)}_{G_N}  (m,a_i) =  \oint \prod_{I=1}^{\ell} {d\phi_I\over 2\pi}   \widetilde Z^{(k)\, gauge}_{G_N} (m,a_i, \phi_I)     \widetilde Z^{(k)\, matter}_{G_N}  (m,a_i, \phi_I)   \,,
 \fe
 where $\ell =k$ for $SU(N), SO(2N)$ and $SO(2N+1)$, while for $USp(2N)$, $\ell =K = \left\lfloor \frac{k}{2} \right\rfloor$. 
The expressions for $\widetilde Z^{(k)\, gauge}_{G_N}$ and  $\widetilde  Z^{(k)\, matter}_{G_N}$  for each group will be summarised below. In the $SU(N)$ case the contour integral was performed explicitly and a general expression for $\partial^2_m\hat Z^{(k)}_{SU(N)}|_{m=0}$ was obtained in \cite{Chester:2019jas} as given in \eqref{kinstsuN}. Therefore this section will focus on other gauge groups. 

A general expression for $\partial^2_m\hat Z^{(k)}_{G_N}|_{m=0}$ is still lacking for gauge groups other than $SU(N)$.  So we will be limited to considering particular examples for these cases.   Below we will present the expressions for $\widetilde  Z^{(k)\, gauge}_{G_N}$  and $\widetilde Z^{(k)\, matter}_{G_N}$ in \eqref{eq:othergroup} for $SO(2N)$, $SO(2N+1)$ and $USp(2N)$, and the prescription of the choice of integration contours, following \cite{Billo:2015pjb, Billo:2015jyt}. From these expressions, explicit results for $\partial^2_m\hat Z^{(k)}_{G_N}|_{m=0}$ are derived and given in section \ref{sec:yminst}, which include the one-instanton results for all classical Lie groups as well as some multiple-instanton examples.\footnote{We would like to thank Francesco Fucito and Francisco Morales for very helpful discussions and for providing their Mathematica code.}  

Below we list the expressions of $\widetilde Z^{(k)\, gauge}_{G_N}$ and  $\widetilde Z^{(k)\, matter}_{G_N}$ in \eqref{eq:othergroup} for the various gauge groups. 

\begin{itemize}

\item $SO(2N)$
  \begin{align} \label{eq:SO2nInt}
\widetilde Z^{(k)\, gauge}_{SO(2N)} (m,a_i, \phi_I) =&\,  {(-1)^k \over 2^k k!}  \left( { \epsilon_+ \over \epsilon_1 \epsilon_2} \right)^k { \Delta(0) \Delta(\epsilon_+) \over  \Delta(\epsilon_1)  \Delta(\epsilon_2) } \prod^k_{I=1} {4 \phi_I (4\phi_I - \epsilon^2_+) \over P(\phi_I +\epsilon_+/2 )P(\phi_I - \epsilon_+/2 )} \, , \\
\widetilde Z^{(k)\, matter}_{SO(2N)}  (m,a_i, \phi_I) =&   \left( { (\epsilon_1+  \epsilon_3)(\epsilon_1 +  \epsilon_4)\over \epsilon_3 \epsilon_4} \right)^k { \Delta(\epsilon_1+  \epsilon_3) \Delta(\epsilon_1+  \epsilon_4) \over  \Delta(\epsilon_3)  \Delta(\epsilon_4) } \cr
& \times \prod^k_{I=1} { P(\phi_I + (\epsilon_3-  \epsilon_4)/2 )P(\phi_I -  (\epsilon_3-  \epsilon_4)/2 ) \over (4\phi_I - \epsilon^2_3) (4\phi_I - \epsilon^2_4) } \, ,  \nonumber
  \end{align}
where $\epsilon_+ =  \epsilon_1+ \epsilon_2$ and $\epsilon_ 3 = m -  \epsilon_+/2, \epsilon_ 4 = -m -  \epsilon_+/2$.  The parameters $\epsilon_1$ and $\epsilon_2$ serve as omega deformations to regulate the instanton partition function.   The functions $P$ and $\Delta$ are defined as
  \ie
 P(x) = \prod_{j=1}^N (x^2 - a^2_j) \, , \qquad  \Delta(x) = \prod_{I<J}^k (x^2 - \phi^2_{IJ}) (x^2 - (\phi^+_{IJ})^2) \, ,
  \fe  
and  $\phi^+_{IJ} = \phi_{I}+ \phi_{J}$.  

The integral is computed by closing the contours in the upper-half complex plan of $\phi_I$, after giving $\epsilon_ i$ an  imaginary part with the following hierarchy \cite{Billo:2015pjb}
\ie \label{eq:eps}
\Im (\epsilon_4) \gg \Im (\epsilon_3) \gg \Im (\epsilon_2) \gg \Im (\epsilon_1) \, . 
\fe
For the case in which the base manifold is $S^4$ that is relevant for our computation of integrated correlators in $\mathcal{N}=4$ SYM, we set $\epsilon_1 = \epsilon_2=1$, but only after the contour integrals are performed using the prescription described above.  This prescription for the choice of contours also applies to the $SO(2N+1)$ and $USp(2N)$ cases that we will discuss next. 
  
\item $SO(2N+1)$
  \begin{align}
\widetilde Z^{(k)\, gauge}_{SO(2N+1)}  (m,a_i, \phi_I) =&\, {(-1)^k \over 2^k k!}  \left( { \epsilon_+ \over \epsilon_1 \epsilon_2} \right)^k { \Delta(0) \Delta(\epsilon_+) \over  \Delta(\epsilon_1)  \Delta(\epsilon_2) } \prod^k_{I=1} {4 \phi_I (4\phi_I - \epsilon^2_+) \over P(\phi_I +\epsilon_+/2 )P(\phi_I - \epsilon_+/2 )}\, , \\
\widetilde Z^{(k)\, matter}_{SO(2N+1)} (m,a_i, \phi_I) =&   \left( { (\epsilon_1+  \epsilon_3)(\epsilon_1+  \epsilon_4)\over \epsilon_3 \epsilon_4} \right)^k { \Delta(\epsilon_1+  \epsilon_3) \Delta(\epsilon_1 +  \epsilon_4) \over  \Delta(\epsilon_3)  \Delta(\epsilon_4) } \cr
&\times  \prod^k_{I=1} { P(\phi_I + (\epsilon_3-  \epsilon_4)/2 )P(\phi_I -  (\epsilon_3-  \epsilon_4)/2 ) \over (4\phi_I - \epsilon^2_3) (4\phi_I - \epsilon^2_4) } \, , \nonumber
  \end{align}
  with 
  \ie
 P(x) =x \prod_{j=1}^N (x^2 - a^2_j) \, , \qquad  \Delta(x) = \prod_{I<J}^k (x^2 - \phi^2_{IJ}) (x^2 - (\phi^+_{IJ})^2) \, ,
  \fe  
and  $\phi^+_{IJ} = \phi_{I}+ \phi_{J}$.

\item $USp(2N)$
  \begin{align}
\widetilde Z^{(k)\, gauge}_{USp(2N)}(m,a_i,& \phi_I)  = {(-1)^k \over 2^k k!}  { \left( \epsilon_+ \right)^{k-\nu} \over \left( \epsilon_1 \epsilon_2 \right)^k } { \Delta(0) \Delta(\epsilon_+) \over  \Delta(\epsilon_1)  \Delta(\epsilon_2) } {1 \over P(\epsilon_+/2)^{\nu}} \prod^K_{I=1} {4 \phi_I (4\phi_I - \epsilon^2_+) \over P(\phi_I +\epsilon_+/2 )P(\phi_I - \epsilon_+/2 )} \, , \\
\widetilde Z^{(k)\, matter}_{USp(2N)} (m,a_i,& \phi_I) =   { \left(  (\epsilon_1+  \epsilon_3)(\epsilon_1+  \epsilon_4) \right)^{k+\nu} \over \left( \epsilon_3 \epsilon_4 \right)^k } { \Delta(\epsilon_1+  \epsilon_3) \Delta(\epsilon_1 +  \epsilon_4) \over  \Delta(\epsilon_3)  \Delta(\epsilon_4) } P((\epsilon_3 -\epsilon_4)/2)^{\nu} \cr
  &\times \prod^K_{I=1} { P(\phi_I + (\epsilon_3-  \epsilon_4)/2 )P(\phi_I -  (\epsilon_3-  \epsilon_4)/2 )  (4\phi_I - (\epsilon_1+\epsilon_3)^2) (4\phi_I - (\epsilon_1+\epsilon_4)^2) } \, , \nn
  \label{USpninst}
  \end{align}
where $k=2K+\nu$ and $\nu=1$ if $k$ is odd, $\nu=0$ if $k$ is even. Furthermore,  the functions $P, \Delta$ are defined as
  \ie
 P(x) =x \prod_{j=1}^N \big(x^2 - {a^2_j}/{2} \big) \, , \qquad  \Delta(x) = \prod_{I<J}^K (x^2 - \phi^2_{IJ}) (x^2 - (\phi^+_{IJ})^2) \prod_{I=1}^K (x^2 - \phi^2_I)^{\nu}\, .
  \fe  
 \end{itemize}

%%%%%
\subsection{One instanton contribution to $\C_{SO(n)}$}
\label{1-instantonN}
%%%%%

Here we consider the large-$y$ expansion, with $y=\pi \tau_2 = 4\pi^2/g_{YM}^2$, of the one-instanton contribution to $\C_{SO(n)}(\tau, \bar \tau)$ for any $n$, i.e. the perturbation expansion in the one-instanton sector. In the large-$y$ expansion, the one-instanton term can be expressed as
\ie \label{eq:1-inst}
\left \langle \partial_m^2 \hat{Z}^{(1)}_{SO(n)}(m, a_i) \Big{|}_{m=0}  \right \rangle_{SO(n)}  =e^{2\pi i\tau} \left[ Y_0 (N) + Y_1(N) {1\over y}+ Y_{2}(N) {1\over y^2} + \cdots \right]\, ,
\fe
where $n=2N$ or $n=2N+1$, and the one-instanton contribution to the integrated correlator is given by
\ie
\C^{(1)}_{SO(n)}(\tau, \bar \tau) = \tau_2^2 \partial_{\tau} \partial_{\bar \tau} \left \langle \partial_m^2 \hat{Z}^{(1)}_{SO(n)}(m, a_i) \Big{|}_{m=0}  \right \rangle_{SO(n)} \, . 
\fe
The task is to determine the coefficient functions $Y_i (N)$ in \eqref{eq:1-inst}. 

This is done by expanding $\partial_m^2 \hat{Z}^{(1)}_{SO(n)}(m, a_i) \Big{|}_{m=0}$, as given in \eqref{kinstso2N} for $SO(2N)$ and \eqref{kinstso2N1} for $SO(2N+1)$, in the small-$a_i$ expansion for any $N$ (here we have expanded them to order $a_i^4$). We then take the expectation value according to the matrix model integrals given in \eqref{SO1exp} and \eqref{SO2exp} for $SO(2N)$ and $SO(2N+1)$, respectively. We find the coefficients $Y_i (N)$ obey the following recursion relations:  
\begin{align}
& (2 n+3) (2 n+5) (4 n+9) Y_0(N)-\left(160 n^3+696 n^2+728 n+87 \right)
   Y_0(N+1) \cr
   &+36 (n+1) (n+2) (4 n+1) Y_0(N+2) =0 \, , \\
   \cr
  & (n+1) (n+2) (2 n+1) (2 n+3) (4 n+9) Y_1(N)- 5 \left(32 n^4+56 n^3-176 n^2-401 n-183\right)Y_1(N+1)\cr
   & +\, 36 (n-3) n (n+1) (n+2) (4 n+1) Y_1(N+2) =0\, , \\
   \cr
  &  (n+1) (n+2) (2 n-1) (2 n+1) \left(8 n^5-6 n^4-182 n^3-153 n^2+333
   n+270\right)Y_2(N) \cr
   & -  \,  (n-1) n \left(320 n^7-1872 n^6-5984 n^5+25526 n^4+17178 n^3-70475 n^2+1587
   n+46962\right)Y_2(N+1) \cr
   &+ \, 36 (n-1) n (n+1) (n+2) \left(8 n^5-86 n^4+186 n^3+155 n^2-407 n+96\right)Y_2(N+2) =0 \, .
\end{align}
These equations apply to both $SO(2N)$ (i.e. using $n=2N$) and $SO(2N+1)$ (i.e. using $n=2N+1$). Furthermore, the recursion relations can also be solved order by order in $1/n$ expansion, once the initial condition is given. We have used these relations to verify the large-$\fluxSO$ results given in \eqref{eq:obr}. 

\section{Laplace-difference equations}
\label{sec:laplace-dif}

In this appendix we will review the evidence for the Laplace-difference equations that hold for any classical gauge group and are summarised in section  \ref{sec:lapdiff}.  These equations determine  the integrated correlators for any classical gauge group in terms of  $\C_{SU(2)}(\tau, \bar{\tau})$, the integrated correlator for the gauge group $SU(2)$. 

We begin by reviewing the $SU(N)$  Laplace-difference equations, \eqref{lapdiffSUN}, satisfied by $\cC_{SU(N)}(\tau,\bar\tau)$, which are  are  described in more detail in  \cite{Dorigoni:2021bvj,Dorigoni:2021guq}. 
The integrated correlator   with gauge group $SU(N)$ has the form \eqref{CNexp}
  \begin{align}
\cC_{SU(N)} (\tau,\bar\tau)  = {1\over 2}  \sum_{(m,n)\in\mathbb{Z}^2}  \int_0^\infty \exp\Big(- t \pi \frac{|m+n\tau|^2}{\tau_2} \Big) B_{SU(N)}(t) \, dt\,,
\label{gsun}
\end{align} 
which is the same as  \eqref{mainres} with $B^2_{G_N}(t)=0$ and $B_{SU(N)}(t) \equiv B^1_{SU(N)}(t)$.  
The function $B_{SU(N)}(t)$ has the form given by   \eqref{eq:BSUN}  
 \bea 
 B_{SU(N)} (t)=\frac{\cQ_{SU(N)} (t)}{(t+1)^{2N+1}}\,,
 \label{bndef}
 \eea
 where  $\cQ_{SU(N)} (t)$ is a polynomial of degree $2N-1$ given by \eqref{polydef}.
 In applying  the Laplace operator to $\cC_{SU(N)}(\tau,\bar\tau)$ we  note the important relation
 \bea
\Delta_\tau e^{-t\pi Y(\tau,\bar\tau) }= e^{-t\pi Y(\tau,\bar\tau) }\left[ (\pi t Y(\tau,\bar\tau) )^2-2\pi t Y(\tau,\bar\tau) \right] =  t\,   \partial_t^2    \left( t\, e^{-t \pi Y(\tau,\bar\tau) }\right) \,,
\label{lapint} 
\eea
where
\bea
Y(\tau,\bar\tau) =   \frac{|m+n\tau|^2}{\tau_2} \,.
\label{taut}
\eea
It therefore follows that applying $\Delta_\tau$ to \eqref{gsun} and after integration by parts, we obtain 
\bea
\Delta_\tau\cC_{SU(N)}(\tau,\bar\tau) = \frac{1}{2} \sum_{(m,n)\in \Z^2}\int_0^\infty e^{-t\pi\frac{|m+n \tau|^2}{\tau_2} } \,t\, \frac{d^2}  {dt^2} \Big[ t\, B_{SU(N)}(t)\Big]\,dt\, .
\label{lapong}
\eea

To proceed, we note that Jacobi polynomials satisfy the following three-term recursion relation
 \bea \label{bndef} 
 &\notag  2(n+\alpha-1)(n+\beta-1)(2n+\alpha+\beta)P_{n-2}^{(\alpha,\beta)}(z)+ 2n(n+\alpha+\beta)(2n+\alpha+\beta-2) P_n^{(\alpha,\beta)}(z)\\
 &= (2n+\alpha+\beta-1)\Big[(2n+\alpha+\beta)(2n+\alpha+\beta-2)z +\alpha^2 -\beta^2\Big]P_{n-1}^{(\alpha,\beta)}(z)\,.
 \eea
as well as
\bea
(z-1)\frac{d}{dz}P_n^{(\alpha,\beta)}(z) = n\, P_{n}^{(\alpha, \beta) }(z) - (\alpha+n)\,   P_{n-1}^{(\alpha,\beta+1) }(z) \,.
\label{diffjac}
\eea
From the definition of $B_{SU(N)}(t)$ we find
\ie
t\, \frac{d^2}  {dt^2} \Big[ t\, B_{SU(N)}(t)\Big]-  4c_{SU(N)} & \left[  B_{SU(N+1)}(t) - 2 B_{SU(N)}(t) + B_{SU(N-1)}(t) \right] \cr
& - (N+1) B_{SU(N+1)}(t) - (N-1) B_{SU(N+1)}(t) = 0 \,.
\label{brecur}
\fe
 Substituting this relation into \eqref{lapong} gives the Laplace-difference equation \eqref{lapdiffSUN},

We now turn to the Laplace-difference equations for the integrated correlators of theories with the other general classical gauge groups \eqref{lapdiffSO} and \eqref{lapdiffUSp}. Once again the equations are equivalent to differential-difference equations for the rational functions $B^i_{G_N}(t)$ given in subsection \ref{sec:exact-exp}, namely \eqref{eq:SONnew} for  $B^{i}_{SO(2N)}(t)$ as well as \eqref{BSOodd2def}, \eqref{eq:B2new} and \eqref{eq:B1new} for $B^{i}_{SO(2N+1)}(t)$ and, equivalently, $B^{i}_{USp(2N)}(t)$.

  In the case of $SO(n)$ gauge groups the differential recurrence relation is 
  \begin{align}
t\, \frac{d^2}  {dt^2} \Big[ t B^i_{SO(n)}(t) \Big]  -  2 c_{SO(n)} & \Big[ B^i_{SO(n+2)}(t)  -2 \, B^i_{SO(n)}(t)  + B^i_{SO(n-2)}(t)   \Big] \nn\\
& - n\, B^i_{SU(n-1)}(t) +(n-1)\, B^i_{SU(n)}(t)=0 \, ,
\label{dB-SO}
\end{align} 
while for $USp(n)$ (with $n=2N$) it takes a very similar form, 
\begin{align}
t\, \frac{d^2}  {dt^2} \Big[ t B^i_{USp(n)}(t) \Big]  -  2 c_{USp(n)} & \Big[ B^i_{USp(n+2)}(t)  -2 \, B^i_{USp(n)}(t)  + B^i_{USp(n-2)}(t)   \Big] \nn\\
& + n\, B^{i'}_{SU(n+1)}(t) -(n+1)\, B^{i'}_{SU(n)}(t)=0 \, .
\label{dB-Sp}
\end{align} 
Note that the rescaling $\tau \to 2 \tau, \bar{\tau} \to 2 \bar{\tau}$ in the second line of \eqref{lapdiffUSp} implies that, in the above equation \eqref{dB-Sp}, $B^{i'}_{SU(n)}(t)=B_{SU(n)}(t)$ when $i=2$ (as given in  \eqref{eq:BSUN}), and $B^{i'}_{SU(n)}(t)=0$ when $i=1$. Using explicit expressions for $B^i_{G_N}(t)$ given in subsection \ref{sec:exact-exp}, it is straightforward to verify \eqref{dB-SO} and \eqref{dB-Sp} for any given $N$. Furthermore, the Laplace-difference equations \eqref{lapdiffSO} and \eqref{lapdiffUSp} on the integrated correlators follow from the above equations.

\section{Matching with string theory in  $AdS_5\times S^5/\Z_2$  orientifold}
\label{sec:stringy}

In this appendix we will briefly review the type IIB string theory description that is the holographic dual of the $\cN=4$ SYM theories with classical gauge groups $G_N$.

The holographic equivalence between $\cN=4$ SYM theory and type IIB superstring theory was initially formulated in the context of the $SU(N)$ gauge theory  \cite{Maldacena:1997re, Gubser:1998bc, Witten:1998qj}.  It was argued that in the large-$N$ limit the gauge theory is dual to the string theory in $AdS_5\times S^5$, which is the near horizon geometry of a stack of $N$ $D3$-branes.  According to this correspondence the string coupling is related to the Yang--Mills coupling by $g_s=g_{_{YM}}^2/4\pi$ and the $AdS_5\times S^5$  length scale, $L$,  is related to the RR five-form flux  $N$ by  $(L/\ell_s)^4 =g_{_{YM}}^2 N$ (where $\ell_s$ is the string length scale).   This was soon extended to more general gauge groups and corresponding geometries.  

Of particular relevance is the generalisation to  theories with general classical gauge groups  that still preserve maximal supersymmetry \cite{Elitzur:1998ju, Witten:1998xy}. These are type IIB string theories in an  orientifold with background $AdS_5\times (S^5/\Z_2)\sim AdS_5\times RP^5 $. Such backgrounds emerge from the near horizon geometry of $N$ coincident parallel $D3$-branes that are coincident with a parallel orientifold 3-plane ($O3$-plane).   This is the  fixed plane of the orientifold projection $\Omega$, which acts on the string world-sheet and the Chan-Paton factors.     The fact that the action of $\Z_2$ on $S^5$ is free means that there are  no open strings in the type IIB theory in this background and there are also no winding closed strings.   
The orientifold projection leads to non-orientable string world-sheet contributions in the large-$N$ string perturbation theory obtained from $SO(2N)$, $SO(2N+1)$ and $USp(2N)$ $\cN=4$ SYM.     There are four varieties of $O3$-planes that are distinguished by their discrete torsion   \cite{Witten:1998xy}. This  means that they are distinguished by their couplings to $B_{\rm{NSNS}}$ and $B_{\rm{RR}}$ (the  Neveu--Schwarz/ Neveu--Schwarz  and Ramond-Ramond two-form potentials), which are flat connections, i.e. $H= dB=0$, in order to preserve supersymmetry.  

The functional integral over a world-sheet $\Sigma$ includes the phase factors 
$e^{2\pi i \int_\Sigma B_{\rm{NSNS}}}=  e^{2\pi i\theta_{\rm{NSNS}}}$,  and
$e^{ 2\pi i \int_\Sigma B_{\rm{RR}}} =  e^{2 \pi i \theta_{\rm{RR}}}$
where the torsions $\theta_{\rm{NSNS}}$, $\theta_{\rm{RR}}$ take the values $0$, $\frac{1}{2}$, 
and transform as a doublet under $SL(2,\Z)$. The various combinations of $O3$-planes that arise are interpreted as follows:
\begin{itemize}

\item
The orientifold plane with  $(\theta_{\rm{NSNS}}, \theta_{\rm{RR}}) = (0,0)$ is commonly called the  $O3^-$-plane.  It  is $SL(2,\Z)$-invariant and corresponds to the $SO(2N) $ theory.  This plane carries $-\frac{1}{4}$ units of five-form  RR flux.  Together with the $N$ $D3$-branes and their mirror images  the total flux  of this background is $\tilde{N}_{SO(2N)} = (N-\frac{1}{4})$.   

\item
The other three possible combinations are transformed into each other by $SL(2,\Z)$ \cite{Witten:1998xy}.  The $(0, \frac{1}{2})$ case is the $\tilde O3^-$ plane.  This is  invariant under the self-duality group $\Gamma_0(2)$ and corresponds to the $SO(2N+1)$ theory. The $\tilde O3^-$ plane carries {$-\frac{1}{4}$} units of flux.  However one $D3$-brane is necessarily stuck to it since it coincides with its mirror image.  Such a stuck $D3$-brane carries $+\frac{1}{2}$ units of RR flux.  Together with the flux of the remaining $N$ $D3$-branes and their mirror images, the total flux in this background is  $\tilde{N}_{SO(2N+1)}=(N+\frac{1}{4})$.

\item
The $(\frac{1}{2},0)$ and $(\frac{1}{2},\frac{1}{2})$ cases are known as $O3^+$ and $\tilde O3^+$, respectively.   These correspond to  the $USp(2N)$ theory in two different duality frames.  They are transformed into each other by $\Gamma_0(2)$, which interchanges the monopole states and dyonic states  \cite{Hanany:2000fq}.
Since the $O3^+$-plane is  a source of $+\frac{1}{4}$ units of  RR flux, the total flux in the presence of $N$ $D3$-branes is   $\tilde{N}_{USp(2N)}=(N+\frac{1}{4})$.

\end{itemize}

The relation between the parameters of $\cN=4$ SYM with a general classical gauge group and the length scale in the  holographic dual $AdS_5 \times S^5/\Z_2$  is dependent on the RR flux, $\flux$,  of the background. This relation
was motivated in  \cite{Blau:1999vz} by matching the expressions for the trace anomaly in the gauge theory and its holographic supergravity dual.    resulting in  $(L/\ell_s)^4= g_{_{YM}}^2\,\flux$.  This generalises the $SU(N)$ gauge theory result and accounts for the values of the strong coupling parameters given in \eqref{eq:lamval}

 In  the absence of an orientifold projection (i.e. in the large-$N$ $SU(N)$  gauge theory) the world-sheets of string perturbation theory are orientable.  The lowest order contribution arises from a spherical world-sheet of order $1/g_s^2$ and  the next from a toroidal world-sheet of order $g_s^0$. However,  as emphasised in  \cite{Witten:1998xy, Fiol:2014fla} the orientifold projection results in non-orientable string world-sheets that requires the presence of the cross-cap (an $RP^2$  world-sheet), which is of order $1/g_s$ together with non-orientable world-sheets of higher genus.     
At large $N$ a world-sheet of genus $g$ and $s$ factors of $RP^2$ is of order $N^{2-2g-s}$.  Consequently, the large $N$ expansion is an expansion in powers of $1/N$, rather than $1/N^2$.  Furthermore, the replacement $N\to -N$ gives a factor of $(-1)^s$, which clarifies the stringy description of the connection between $SO(2N)$ and $USp(2N)$ theories noted in  \cite{Mkrtchian:1981bb, Cvitanovic:1982bq}.

\bibliographystyle{ssg}
\bibliography{SOSp-ref}
	
\end{document}